\documentclass[11pt]{article}

\RequirePackage{latexsym}
\RequirePackage{graphicx}
\RequirePackage{amssymb}
\RequirePackage{natbib}
\RequirePackage{setspace}

\usepackage{color}
\usepackage{array}
\usepackage{hyphenat}
\usepackage{fancyvrb}
\usepackage{fixltx2e}

\begin{document}
\definecolor{gold}{rgb}{0.85,0.66,0}
\definecolor{grey}{rgb}{0.5,0.5,0.5}
\definecolor{blue}{rgb}{0,0,1}
\definecolor{red}{rgb}{1,0,0}
\definecolor{green}{rgb}{0,0.7,0}
\definecolor{blue}{rgb}{0,0,0}

     \begin{figure}
     \begin{center}
     \includegraphics[scale=1]{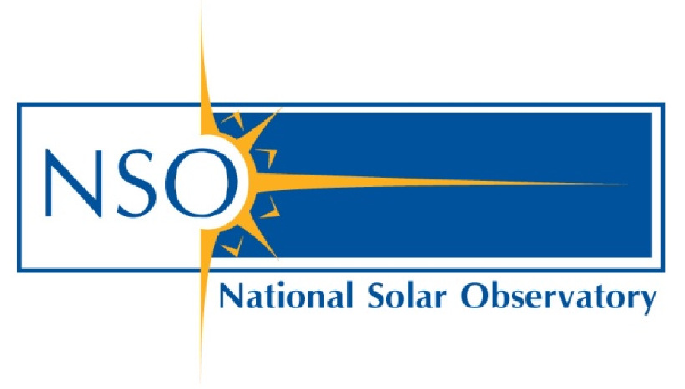}
     \end{center}
     \end{figure}
\title{{\bf GONG ClassicMerge:  Pipeline and Product}
\\$\,$}
\author{Anna L.~H.~Hughes, Kiran Jain, Shukur Kholikov, and the NISP Solar Interior Group
\\$\,$
\\ National Solar Observatory
\\$\,$  \\$\,$  \\$\,$}
\maketitle

\noindent\rule{\linewidth}{0.2mm}
\rule{\linewidth}{0.5mm}

\begin{center}
Technical Report No. {\bf NSO/NISP-2016-001}
\\$\,$
\end{center}

\begin{abstract}

A recent processing effort has been undertaken in order to extend the range-of-coverage of the GONG merged dopplergrams.  The GONG-Classic--era observations have now been merged to provide, albeit at lower resolution, mrvzi data as far back as May of 1995.  The contents of this document provide an overview of what these data look like, the processing steps used to generate them from the original site observations, and the outcomes of a few initial quality-assurance tests designed to validate the final merged images.  Based on these tests, the GONG project is releasing this data product to the user community (http://nisp.nso.edu/data).

\end{abstract}

\pagebreak
\tableofcontents
$\,$

\section[\textcolor{blue}{GONG Overview}]{\textcolor{blue}{GONG Overview}}
\label{GONG}

The Global Oscillation Network Group (GONG) project began taking observations in 1995 with the aim of providing a large, continuous set of solar dopplergram observations for use in helioseismology.  It uses a ground-based network of six sites located around the world and in both hemispheres to acquire observations (weather permitting) 24 hours a day.  The GONG telescopes are of identical design and construction and are located in Learmonth, Australia (LE); Udaipur, India (UD); El Teide in the Canary Islands (TD); Cerro Tololo, Chile (CT); Big Bear, California USA (BB); and Mauna Loa, Hawaii USA (ML).

Originally intended as a three-year project, after multiple upgrades and extensions, GONG remains a valuable source of solar observations today.  In 2001, the sites were each fitted with much higher resolution cameras that have square (rather than the formerly rectangular) pixels.  The two epochs corresponding to the original and upgraded cameras have hence been referred to as GONG-Classic and GONG+, respectively.  Beyond improved spatial resolution, the GONG+ observing strategy was optimized to produce additional data products that are subject to expanded data processing.  An example of the latter is the ImageMerge code, which combines circularly registered GONG dopplergrams and magnetograms taken concurrently at different sites.  The synoptic series of merged dopplergrams, in particular, is broadly useful for global helioseismology studies.  A recent campaign to similarly merge the GONG-Classic dopplergrams was motivated by the desire to extend the time range of this dataset and is the subject matter of this technical report.

Merging GONG dopplergrams requires the coordination of a number of different observation-, calibration-, and intermediate-product types.  Below, we provide a brief primer on the three-letter codes used to identify those types relevant to this product:
\begin{list}{}{}
\item {\bf coy} - Tables of coefficients similar to the {\bf hiy} files, but interpolated between {\bf hiy} sets to provide smoothly varying curves for every GONG site day.
\item {\bf dft} - Calibration images taken with the telescope tracking turned off, allowing the solar disk to drift across the image (providing empirical evidence for the precise east-west line across the image at a given time of day).
\item {\bf hiy} - Tables of coefficients used to provide angles as a function of hour angle that correct for small, smoothly-varying offsets in the GONG images from strict solar-north image alignment.  These alignment coefficients are computed from data compiled over an 8--30 day window, combing {\bf dft} data with cross-correlation angles between images from all six network sites.
\item {\bf qac} - Image statistics data files, containing, e.g., mean-velocity values across the observed solar disks for all observations taken on a given site day.
\item {\bf vzi} - Standard, fully calibrated GONG dopplergrams.  Merged dopplergrams are distinguished from individual site doppergrams with the `site' code: `MR', yielding a 5-letter observation prefix of {\bf mrvzi} (as opposed to, e.g., bb{\bf vzi}).
\end{list}

\section[\textcolor{blue}{ClassicMerge: Basic Product}]{\textcolor{blue}{ClassicMerge: Basic Product}}
\label{BasicProduct}

\begin{figure*}[h]
     \begin{center}
     \includegraphics[scale=0.20]{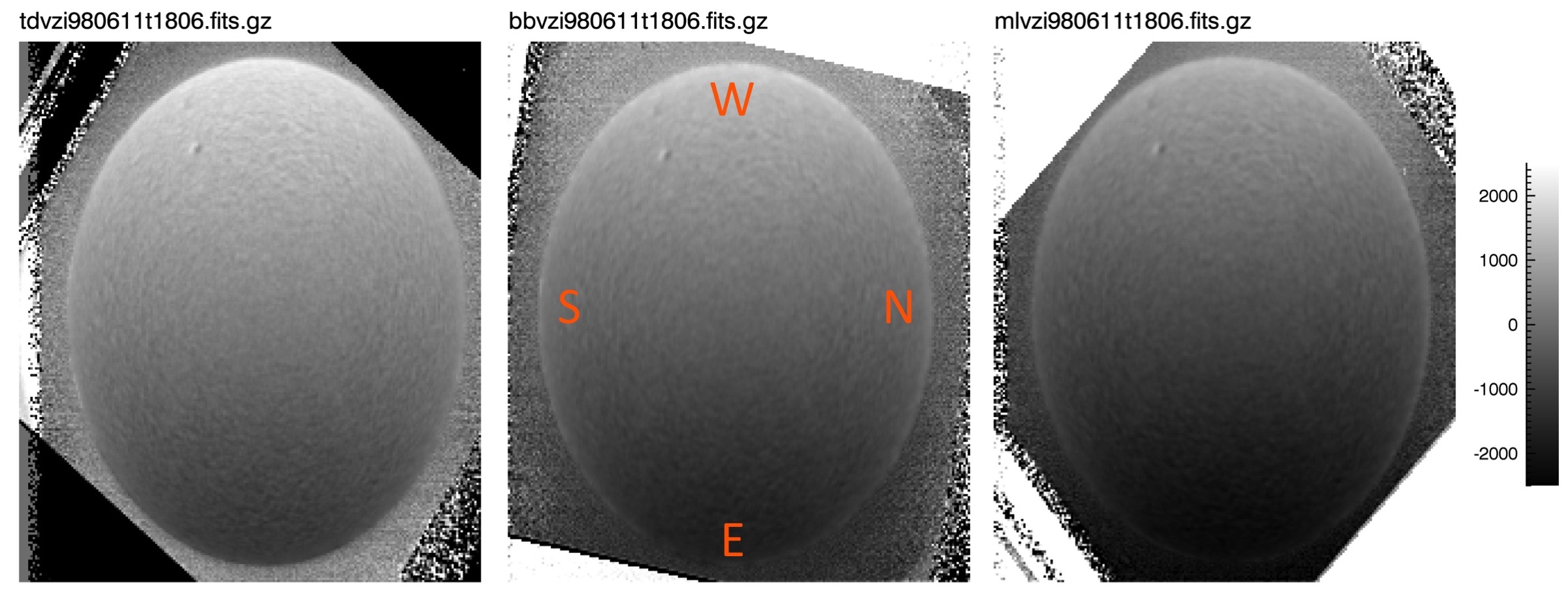}
     \caption[Example GONG-Classic input dopplergrams]{Input GONG-Classic site data used to create the merged output in Figure~\ref{FIG_BP_mergedoutput}.}
     \label{FIG_BP_siteinputset}
     \end{center}
     \end{figure*}

Archived solar dopplergrams created during the GONG-Classic epoch of the GONG program (1995--2001) were processed with the ClassicMerge pipeline.  The GONG-Classic cameras had rectangular pixels (aspect ratio of 1.28:1) oriented with the longer (lower resolution) dimension aligned with the solar axis of rotation.  The resulting images were 204 x 239 pixels across (for a pixel resolution of $\sim$10 x 8 arcseconds), as shown in Figure~\ref{FIG_BP_siteinputset}.
     
\begin{figure}
     \begin{center}
     \includegraphics[scale=0.13]{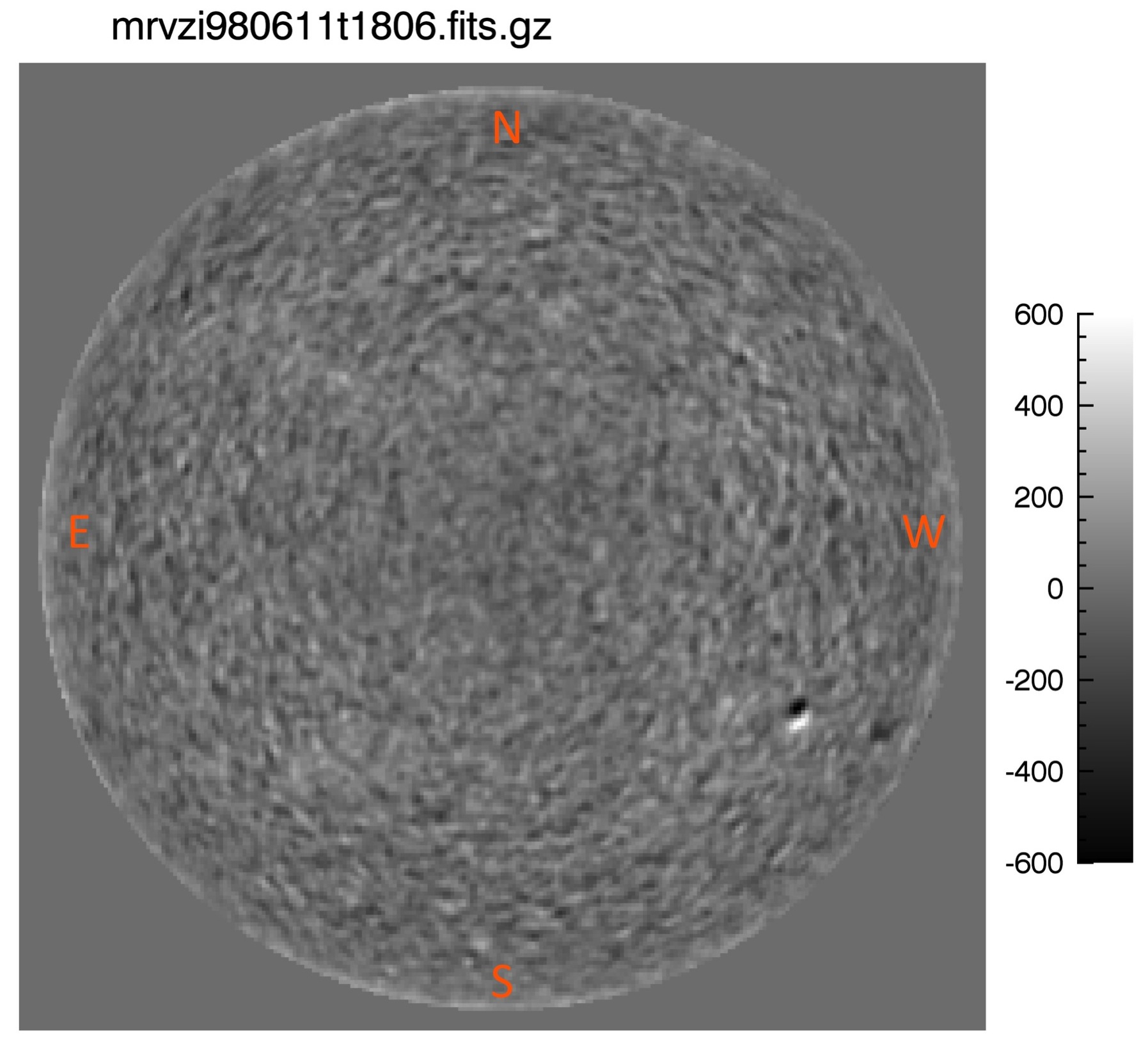}
     \caption[Example merged dopplergram]{Example of a GONG ClassicMerge output dopplergram: 1998-06-11 at 18:06 UT.}
     \label{FIG_BP_mergedoutput}
     \end{center}
     \end{figure}

As in the GONG+ pipeline, during the image-merge processing, individual site observations are registered onto images with a circularized disk of fixed radius before being combined.  To roughly match the resolution of the input images, GONG ClassicMerge images are set to a solar radius of 120 pixels, for an output image size of 251 x 251 pixels, as shown in Figure~\ref{FIG_BP_mergedoutput}.

However, the merge process requires more than just the site observations by themselves, as the observations need to be calibrated and rotationally aligned before they can be merged.  Therefore, the GONG ClassicMerge pipeline pulls additional data from the archives besides the observations, specifically: 
\begin{list}{-}{}
   \item Archived {\bf qac} files listing the mean velocities across the solar disk were ingested in order to (re)compute VELSCALE and VEL\_BIAS, keywords which are used to correct observed-velocity curves (and therefore individual dopplergrams) to match the known ephemeris curves.
   \item Archived {\bf coy} files used to provide daily coefficients for fine-scale rotational alignment between different network images \citep{citeTonerHarvey} were ingested to compute the rotational-shift measurement processed through the image-merge during image circularization.
\end{list}
While the {\bf qac} and {\bf coy} data could be recomputed using primarily the site observations themselves, the effort required to align such a reprocessing to the current GONG+ pipelines was deemed outside the scope of this project.  Instead, the archived measurements were taken as given.

\begin{figure}
     \begin{center}
     \includegraphics[scale=0.085]{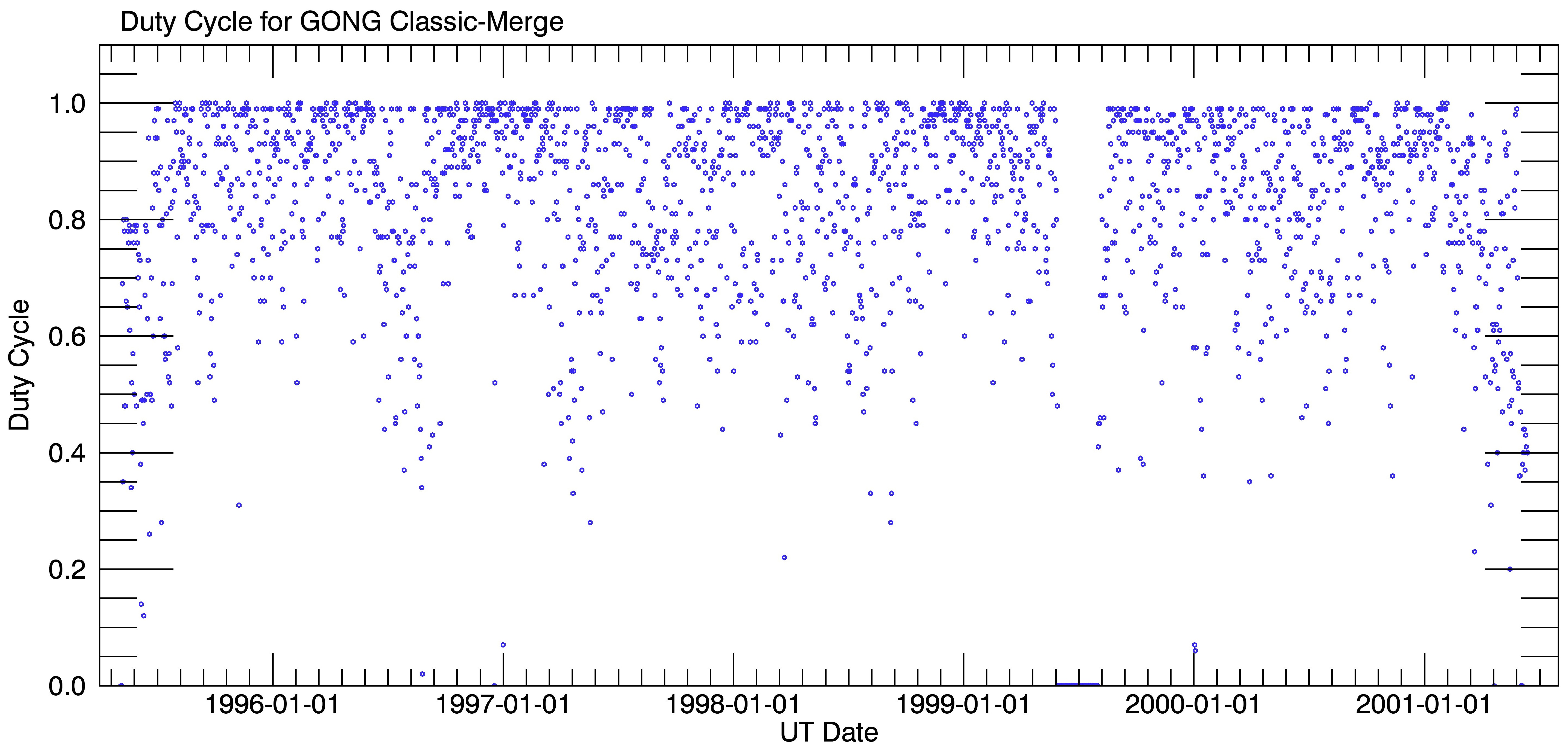}
     \caption[ClassicMerge Duty Cycle]{Duty cycle for the merged GONG-Classic data.}
     \label{FIG_BP_dutycycle}
     \end{center}
     \end{figure}

The final repository of merged dopplergrams produced by the GONG ClassicMerge pipeline covers nearly the full GONG-Classic timespan from May 7th, 1995 to June 13th, 2001, with a median duty cycle of 88\%.  The fractional duty cycle for each day covered is given in Figure~\ref{FIG_BP_dutycycle}.

\section[\textcolor{blue}{Processing Code}]{\textcolor{blue}{Processing Code}}
\label{Pipeline}

This section presents the outline of the pipeline code operations in \S\ref{Pipeline_Layout}, then provides descriptions of specific functions in \S\ref{Pipeline_Functions}.

	\subsection[\textcolor{blue}{Code Layout}]{\textcolor{blue}{Code Layout}}
	\label{Pipeline_Layout}

The flow map for the GONG ClassicMerge pipeline is as follows:
\begin{list}{}{}
\item {\bf gong\_classicmerge\_BatchRun.sh}  {\em START STOP}
	\begin{list}{{\bf-}}{}
	\item Clears the working directory of all files between {\em START} - 2 days and {\em STOP} + 2 days.
	\item For each day from {\em START} to {\em STOP}:
		\begin{list}{$\bullet$}{}
		\item Calls {\bf gong\_classicmerge\_employ.sh} {\em DATE}
			\begin{list}{{\bf $\,\,\, \rightarrow$}}{}
			\item {\bf coy}: For +/- 2 days around {\em DATE}, takes {\bf coy} files from the GONG-Classic archive and copies them into the keep.
			\item {\bf qac}: For +/- 1 day around {\em DATE}, takes {\bf qac} files from the GONG-Classic archive, copies them into the working directories, and unpacks them.
			\item  {\bf vzi}: For +/- 1 day around {\em DATE}, {\bf if} the working directory does not already contain {\bf vzi} files:
				\begin{list}{$\,\, \circ$}{}
				\item Takes the {\bf vzi} files from the GONG-Classic archive, copies them into the working directory, and unzips them.
				\item Deletes files that won't unzip or that have unreadable headers.
				\item Calls {\bf update\_pangle2\_classicmerge.sh} to add OFFSET to the headers.
				\item Calls {\bf gong\_classicmerge\_headeradd.cl} to add ELEV to the headers.
				\item Calls {\bf gong\_classicmerge\_ha\_scalecheck.cl} to assure HA in the headers is listed in the range of $-\pi$ to $+\pi$.
				\item Calls {\bf gong\_classicmerge\_reject} for each retrieved {\bf vzi} file and deletes the ones flagged as to-be-rejected.
				\end{list} 
			\end{list}
		\item Calls {\bf gong\_classicmerge\_employVEL.sh} {\em DATE}
			\begin{list}{{\bf $\,\,\, \rightarrow$}}{}
			\item Site-Day Lists: For the day-of and the day-before {\em DATE} (BB, CT, \& ML) or the day-of and the day-after (LE, TD, \& UD), {\bf if} a list is not already present in the working directory:
				\begin{list}{$\,\, \circ$}{}
				\item Calls {\bf gong\_classicmerge\_sdlist.sh} to output a list of files belonging to that site day.
				\end{list}
			\item {\bf vzi}: For the day-of and the day-before {\em DATE} (BB, CT, \& ML) or the day-of and the day-after (LE, TD, \& UD), {\bf if} a site-day list of files is available:
				\begin{list}{$\,\, \circ$}{}
				\item Checks whether the appropriate {\bf qac} file is also available; if not, deletes the site-day list and all of the {\bf vzi} files it listed (with an exception discussed in \S\ref{DataChecks_QAC}) .
				\item Calls {\bf EPHEMINTERP} to add VCOR[1/2/3/4] to the headers.
				\item Calls {\bf velfit.cl} to add VELSCALE and VEL\_BIAS to the headers.
				\end{list}
			\end{list}
		\item Calls {\bf gong\_classicmerge\_immerge.sh} {\em DATE}
			\begin{list}{$\,\, \circ$}{}
			\item Creates a list of {\bf vzi} files for UT-day {\em DATE}. 
			\item Calls {\bf IMMERGE} to merge {\bf vzi}'s.
			\item Calls {\bf FITSWASH} to clean-up the merged headers.
			\end{list}
		\end{list}
	\item For dates 4 days older than the current IMMERGE processing:
		\begin{list}{$\bullet$}{}
		\item Calls {\bf gong\_classicmerge\_finishcompress.sh} {\em DATE}
			\begin{list}{{\bf $\,\,\, \rightarrow$}}{}
			\item {\bf qac}: Re-compresses the {\bf qac} files into a tarball and moves them from the working directories to the output directories.
			\item {\bf vzi}: For each site (including MR):
				\begin{list}{$\,\, \circ$}{}
				\item Calls {\bf get\_parm} to record all of the updated-header keyword-values to a set of files: OFFSET, ELEV, VCOR[1/2/3/4], VELSCALE, VEL\_BIAS, \& NSITES (N\_IMGMRG).
				\item Re-gzips the {\bf vzi} files and moves them from the working directories to the output directories.
				\end{list}
			\end{list}
		\end{list}
	\item Finishes up by cleaning out any partially-processed data from the working directory.
	\end{list}
\item 
\item {\bf gong\_classicmerge\_dutycycle.sh}
	\begin{list}{{\bf-}}{}
	\item Loops backwards in time over the full span of GONG-Classic-merge dates: 
		\begin{list}{$\,\, \circ$}{}
		\item 2001-06-13 $\rightarrow$ 1995-05-05
		\end{list}
	\item For each date it:
		\begin{list}{$\,\, \circ$}{}
		\item Counts the number of entries where N\_IMGMRG $\ne$ 0 in the GCM\_parm\_mr\_NSITES.dat file of the mrvzi data.
		\item Compares that to 1440 entries for a 100\% duty cycle.
		\item Outputs the computed duty cycle to GCM\_duty\_cycle.dat.
		\end{list}
	\end{list}
\end{list}

	\subsection[\textcolor{blue}{Description of Functions}]{\textcolor{blue}{Description of Functions}}
	\label{Pipeline_Functions}
	
For functions named in the code layout in \S\ref{Pipeline_Layout} above, function descriptions are provided below.

		\subsubsection[\textcolor{blue}{IMMERGE}]{\textcolor{blue}{IMMERGE}}
		\label{Pipeline_Functions_IMMERGE}

This is the same IMMERGE as in the GONG+ pipeline \citep{citeImmerge}, which has recently been updated to make a slight improvement to the ellipse-to-circle image registration.  

The basics of the IMMERGE image transformations are illustrated in the exaggerated Figures~\ref{FIG_PFI_GONGPlusInput}--\ref{FIG_PFI_GONGClassicOutput} below.
These graphics use input images with the same resolution and pixel geometry as standard GONG+ (Figures~\ref{FIG_PFI_GONGPlusInput}$\rightarrow$\ref{FIG_PFI_GONGPlusOutput}) or GONG-Classic (Figures~\ref{FIG_PFI_GONGClassicInput}$\rightarrow$\ref{FIG_PFI_GONGClassicOutput}) data, but the ellipticity and tilt of the solar disk have been artificially increased to emphasize  the nature of the transformations.  Grids have also been overlain on the images to aid in assessing the nature of the IMMERGE-registration transforms, which include:
\begin{enumerate}
\item Flip the image about the lower-left--to--upper-right diagonal to convert from the orientation of the site images (solar-north at right, east-limb at bottom) to the more conventional orientation of the merged images (solar-north at top, east-limb at left).  {\bf Note:} The colored circles in the figures match grid-limb crossing points between the input and output images.
\item Scale and bias correct the site-observation velocities using VELSCALE, VEL\_BIAS, and VCOR1 to account for the Earth's motion and instrumental seeing effects.  {\bf Note:} The TD image used in the GONG+ example has a negative VELSCALE value, which is why the grey-scaling appears reversed between Figure~\ref{FIG_PFI_GONGPlusInput} (site-input) and Figure~\ref{FIG_PFI_GONGPlusOutput} (IMMERGE output).
\item Isolate and remove the rotational velocity of the Sun by subtracting a surface fit to the dopplergram.
\item Interpolate to circularize the originally elliptical solar disk.  In the GONG+ images, the ellipticity is ordinarily very slight and is due primarily to atmospheric refraction.  In the GONG-Classic images, the ellipticity is ordinarily pronounced (but only along the input y-axis) and is due primarily to the rectangular shape of the GONG-Classic camera pixels.
\end{enumerate}
Once these image transformations have been completed, IMMERGE takes the average of each registered-pixel measurement between all observations taken from different GONG sites at a given time and reports these averaged images as the merged mrvzi-file results.

     \begin{figure}
     \begin{center}
     \includegraphics[scale=0.15]{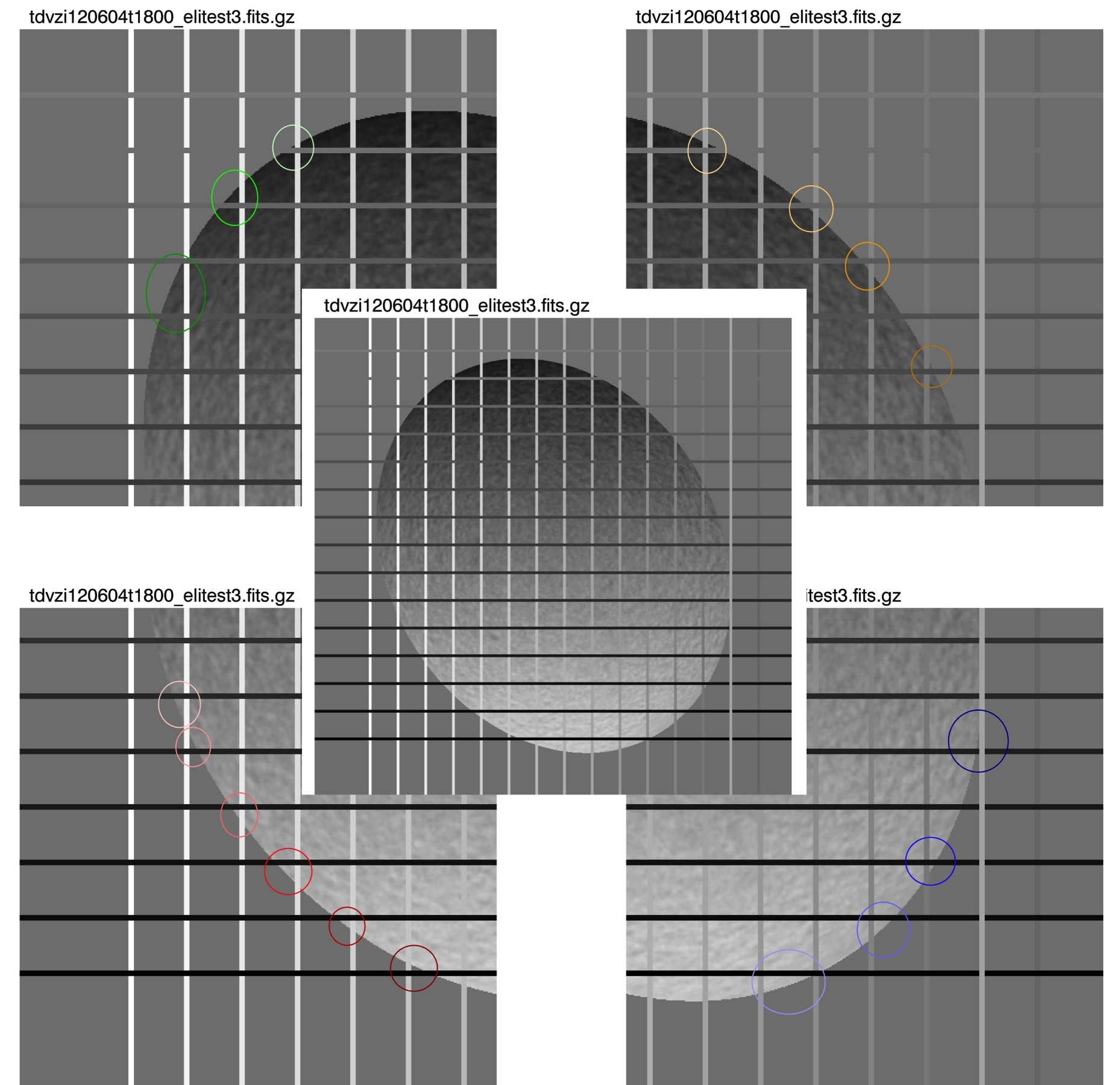}
     \caption[IMMERGE ellipse-test: GONG+ Input image]{GONG+ input-image to IMMERGE for the exaggerated-ellipse test.}
     \label{FIG_PFI_GONGPlusInput}
     \end{center}
     \end{figure}
     \begin{figure}
     \begin{center}
     \includegraphics[scale=0.15]{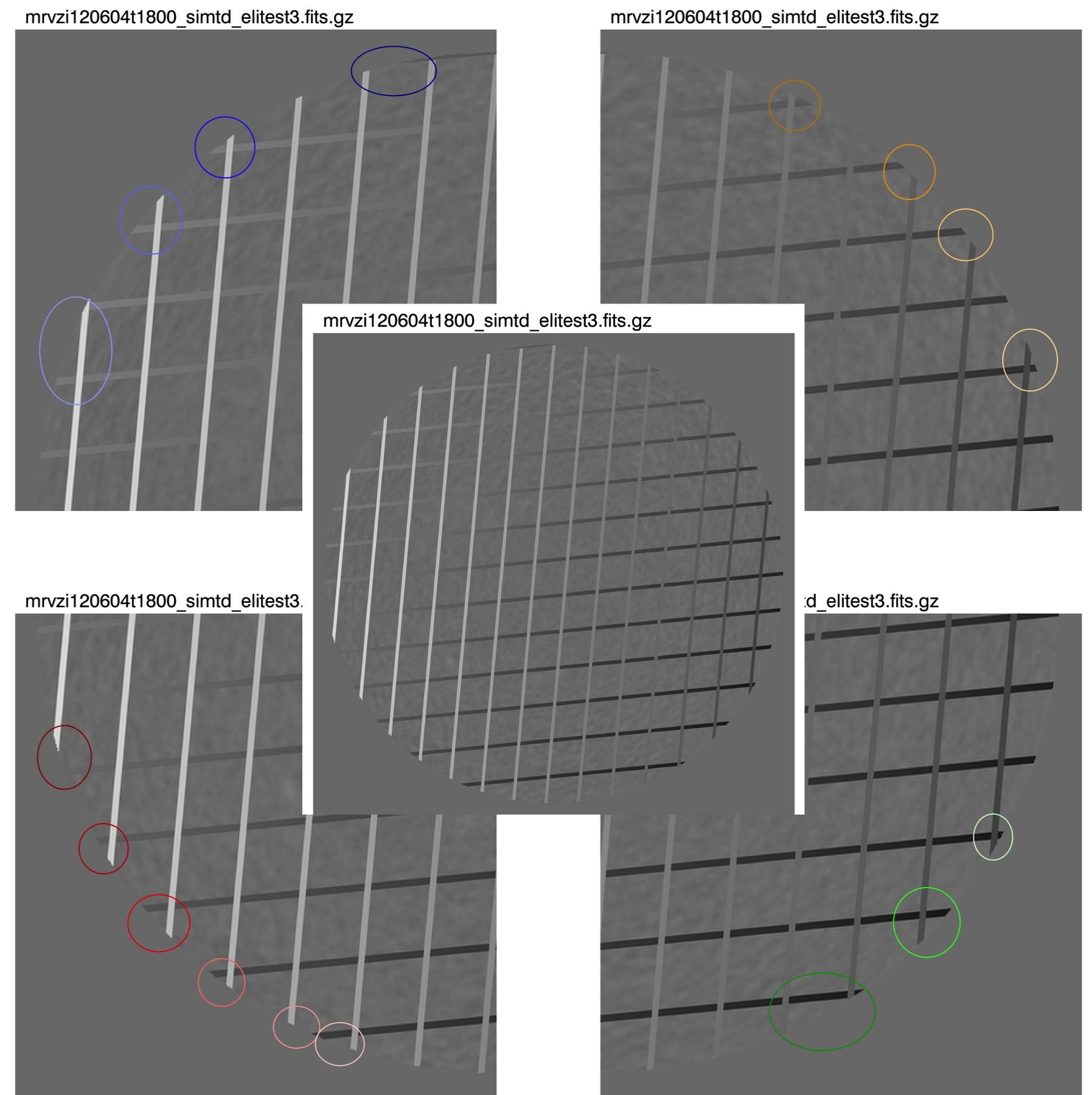}
     \caption[IMMERGE ellipse-test: GONG+ Output image]{GONG+ output image from IMMERGE for the exaggerated-ellipse test.}
     \label{FIG_PFI_GONGPlusOutput}
     \end{center}
     \end{figure}
     \begin{figure}
     \begin{center}
     \includegraphics[scale=0.15]{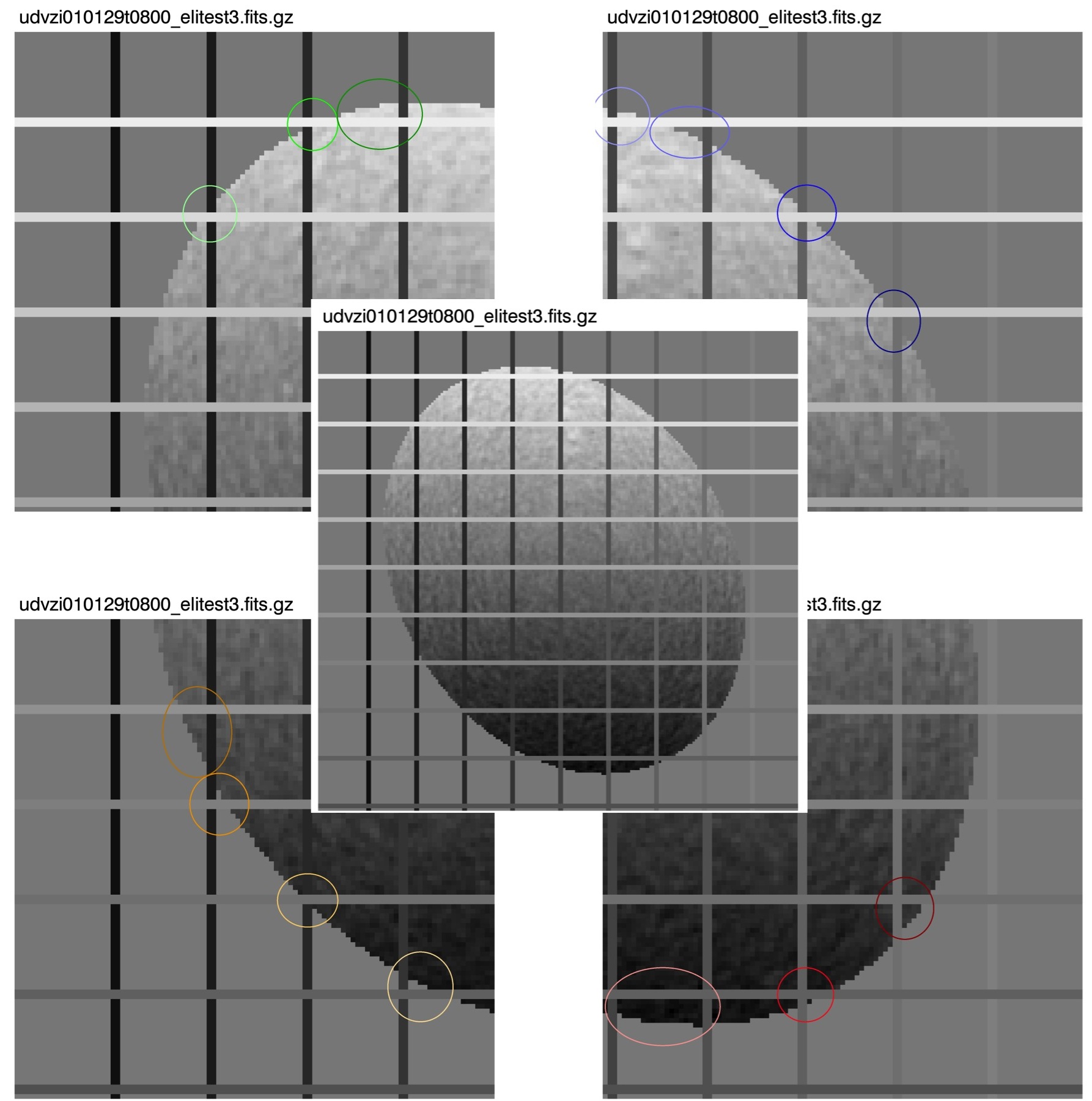}
     \caption[IMMERGE ellipse test: GONG-Classic Input image]{GONG-Classic input-image to IMMERGE for the exaggerated-ellipse test.}
     \label{FIG_PFI_GONGClassicInput}
     \end{center}
     \end{figure}
     \begin{figure}
     \begin{center}
     \includegraphics[scale=0.15]{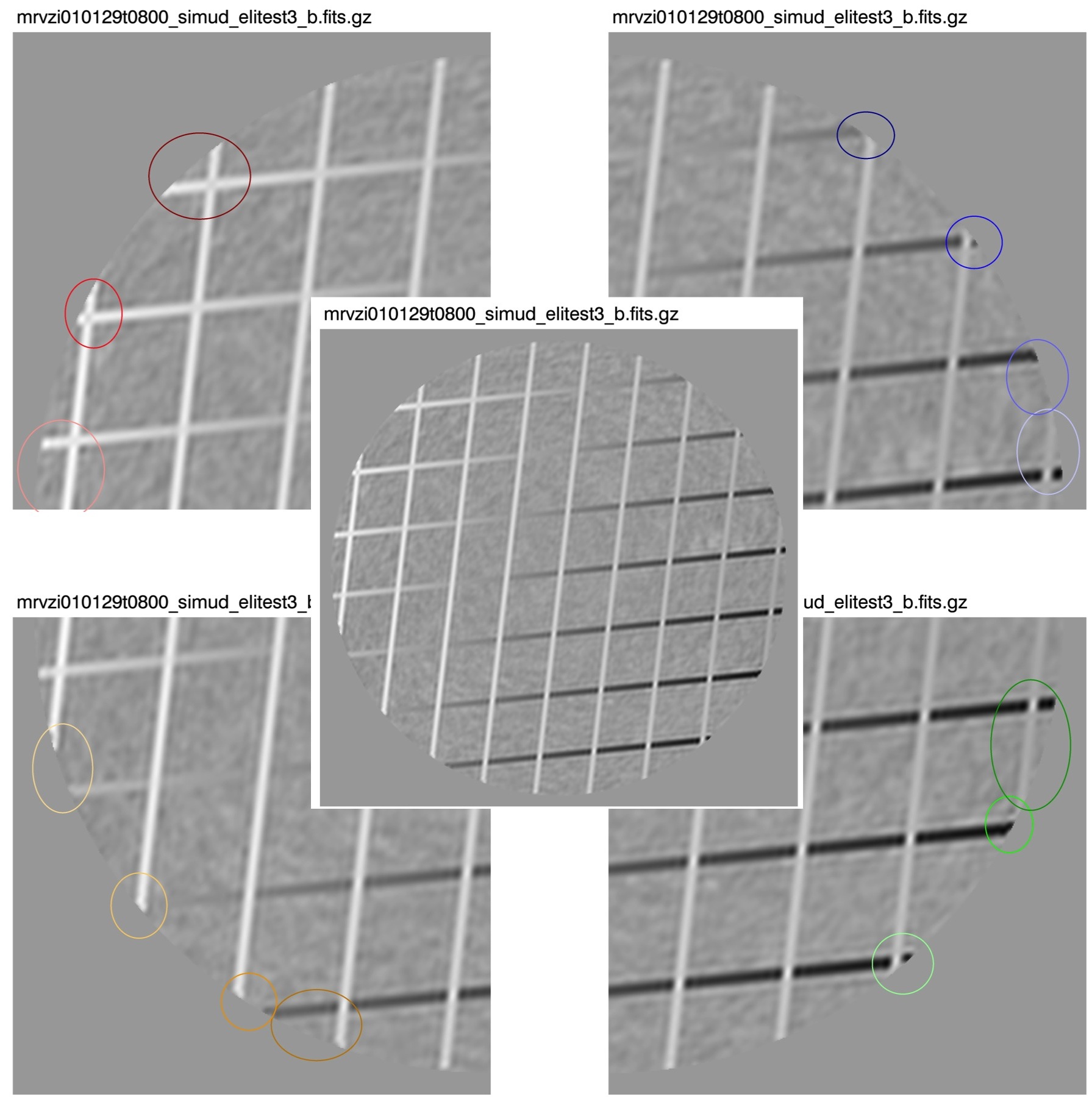}
     \caption[IMMERGE ellipse  test: GONG-Classic Output image]{GONG-Classic output image from IMMERGE for the exaggerated-ellipse test.}
     \label{FIG_PFI_GONGClassicOutput}
     \end{center}
     \end{figure}

The call to IMMERGE from the GONG ClassicMerge pipeline specifies a smaller output image size (251$\times$251 pixels for a 120-pixel radius solar image) than for the merged GONG+ images (400-pixel radius), but otherwise uses the same input requirements and transform algorithms.  Since some of those inputs include the keywords VELSCALE, VCOR1, and OFFSET (the latter being used to indicate the precise orientation of solar-north on the site-images) --- and because much of the GONG-Classic data did not automatically include these keyword values in the image headers --- upstream portions of the GONG ClassicMerge pipeline are set up to ensure that these keyword values get added to the GONG-Classic input files before IMMERGE is called.

		\subsubsection[\textcolor{blue}{gong\_classicmerge\_reject}]{\textcolor{blue}{gong\_classicmerge\_reject}}
		\label{Pipeline_Functions_reject}

This is the primary code for discarding bad GONG-Classic site images from the ClassicMerge pipeline, and it is intended to roughly replicate the functionality of the A.I.R.~(Automated Image Rejection) code \citep{citeAIR}, which operates on the GONG+ pipeline.

All of the checks {\bf gong\_classicmerge\_reject} employs are listed below, and all operate solely based on keyword-values in the FITS-image headers:
\begin{enumerate}
\item Reject files that have the FILLED keyword in their headers (i.e., those observations already flagged for exclusion during previous processing).
\item Reject images with bad pointing.  This is done using the keywords FCOL, LCOL, FROW, and LROW (which denote the first and last usable rows and columns of each image), and FNDLMBXC, FNDLMBYC, FNDLMBMI, FNDLMBMA, C\_MI, and C\_MA (which define the elliptical limb as returned by the FNDLMB and HGEOM routines).  The horizontal and vertical pointing are checked separately, using the following requirements for acceptable pointing:
	\begin{list}{$\bullet$}{}
	\item FNDLMBXC $\pm$ max(FNDLMBMI, C\_MI)  is within the bounds [FCOL -- LCOL]
		\begin{list}{}{}
		\item i.e., $\,\,\, x_\mathrm{center} \pm \max\left(r_\mathrm{minor}\right)$ falls within [FCOL -- LCOL]
		\end{list}
	\item FNDLMBYC $\pm$ max(FNDLMBMA, C\_MA)  is within the bounds  [FROW -- LROW]
		\begin{list}{}{}
		\item i.e., $\,\,\, y_\mathrm{center} \pm \max\left(r_\mathrm{major}\right)$ falls within [FROW -- LROW]
		\end{list}
	\end{list}
This check relies on the assumption that GONG-Classic images, due to their rectangular pixels, naturally have their ellipse-major-axis closely aligned with the y-axis of the input image, and their minor-axis along the x-axis of the image.
\item Reject images whose instrument-reported rotation deviates too far from the expected image rotation. The instrument-recorded rotation is given in each image header by the ROTATOR keyword, whereas the expected rotation, based on the pointing of the telescope, is given by:
	\begin{list}{}{}
	\item $\mathrm{ROT_{expected}} = $ ROLL - PITCH - BETA - PA $\,$.
	\end{list}
Here, ROLL and PITCH are the roll and pitch of the telescope, BETA is the parallactic angle, and PA is the position angle of the Sun.   In the GONG+ A.I.R. code, the maximum allowed deviation is 0.005 radians.  However, to allow for a bit more uncertainty in the GONG-Classic data, the {\bf gong\_classicmerge\_reject} code uses a threshold of 0.01 radians.  
\end{enumerate}

The A.I.R.~code for the GONG+ pipeline also compares the statistical properties of the solar disk pixels against empirically determined limits, as well as with those of the adjacent observations.
{\em However}, as the GONG-Classic dataset is somewhat more limited than the GONG+ data, these checks are currently considered to be outside the scope of the {\bf gong\_classicmerge\_reject} functionality, and are not included in that code.


		\subsubsection[\textcolor{blue}{velfit.cl}]{\textcolor{blue}{velfit.cl}}
		\label{Pipeline_Functions_velfit}

{\bf velfit.cl} is a script that is available within the GRASP package of IRAF (GRASP being one of the primary packages used for processing GONG data).  {\bf velfit.cl} computes and adds the keywords VELSCALE and VEL\_BIAS to GONG images using the same algorithm used in the QA stage of the pipeline VMBICAL processing of the GONG+ data (although the {\bf velfit.cl} script itself is not used in the GONG+ pipeline).  (Please see Appendix~\ref{App_VELFIT} for basic documentation.)

As input, {\bf velfit.cl} requires a list of observations taken on a given site day (i.e., observations from sunup to sundown for one site for one day) as well as a table of mean on-solar-disk velocity values for each of those observations.  The script collects time stamps, hour angles, and ephemeris velocities (VCOR1) from the input headers and computes the difference between the disk-mean observed and ephemeris velocities.

Next a straight line is fit to those velocity differences (for observations taken within $\pm$60 degrees of local noon), and the slope and offset of that line are used to determine the scaling factor (VELSCALE) and bias (VEL\_BIAS), respectively.  Those values can then be used to rescale the observed velocity images so that their means match expected ephemeris values (i.e., to correct for instrumental, day-to-day, refraction-calibration variation in the observed velocities).

{\bf Note:}  The tables of mean-velocity values for each site day that the GONG ClassicMerge pipeline supplies to {\bf velfit.cl} are the {\bf qac} files that are taken directly from the GONG-Classic archive.  When production of the {\bf qac} files was originally implemented, velocity averages were taken across the solar disk from 0 to 99\% of the solar radius.  However, beginning in December of 1995, this was adjusted to cover only 0 to 95\% of the solar radius, in order to avoid noise associated with near-limb velocities.  This change was not concurrent for all sites; therefore, for data taken from December of 1995 through March of 1996, the averages used to compute VELSCALE for some sites cover 95\% of the disk, while for others they still cover 99\% of the disk.  The last dates for which 99\% was used are:
$\,\,\,$ UD: 1995-12-18;
$\,\,\,$ ML: 1996-01-20;
$\,\,\,$ LE/TD/BB: 1996-03-19;
$\,\,\,$ CT: 1996-03-26.


		\subsubsection[\textcolor{blue}{EPHEMINTERP}]{\textcolor{blue}{EPHEMINTERP}}
		\label{Pipeline_Functions_ephem}

EPHEMINTERP is an IRAF routine within the GONG package, GRASP.  It uses cubic interpolation of ephemeris tables to determine several important physical quantities that depend on an observation's date, time, and geographic location.  It then writes these values into the FITS header for each observation, including the keywords L0, B0, P\_ANGLE, SEMIDIAM, and VCOR[1-4].  The last are a set of coefficients designed to correct the observed solar velocities for the motion of the observer (i.e., the orbital and rotational motion of the Earth), and VCOR1 in particular is important as it defines (and allows other routines to correct for) the mean relative velocity at the center of the solar disk.  In order to compute these values, it is necessary for the FITS header to already contain the keyword values for DATE-OBS, TIME-OBS, LAT, LON, ELEV, BAROMET, and TEMP\_OUT; these keywords define the base observing conditions for each observation.  (Please see Appendix~\ref{App_EPHEMINTERP} for basic EPHEMINTERP documentation.)

		\subsubsection[\textcolor{blue}{gong\_classicmerge\_sdlist.sh}]{\textcolor{blue}{gong\_classicmerge\_sdlist.sh}}
		\label{Pipeline_Functions_sdlist}

{\bf gong\_classicmerge\_sdlist.sh} is a script that was written for the GONG ClassicMerge pipeline, which searches the ClassicMerge working area and outputs a list of available observations falling on a given site day.  This allows the code to provide appropriate lists for use in EPHEMINTERP and {\bf velfit.cl}.

The primary hurdle that this script must overcome is the fact that GONG observations are named and archived according to their UT time/date of observation, while the GONG sites at Big Bear (BB), Mauna Loa (ML) and Learmonth (LE) regularly have observing runs that cross the UT date boundary.  All of the site-day-centric products (e.g., {\bf qac} files) are named according to their {\em site-local} (rather than UT-based) dates, which means that a site-day list for LE will usually also include observations timestamped with the {\em previous-day} UT-date (in the UT-evening), while site-day lists for BB (often) and ML (usually) will also include observations timestamped with the {\em following-day} UT date (in the UT-morning).

This also has implications for the GONG ClassicMerge pipeline.  Specifically, in order to merge one UT-day's worth of data, data taken either the (UT) day before or the day after (depending on the site) must {\em also} be pulled from the GONG-Classic archive (and processed up through the point of adding in the VELSCALE and VCOR1 keywords) before IMMERGE can be called on the specified UT day.

		\subsubsection[\textcolor{blue}{update\_pangle2\_classicmerge.sh}]{\textcolor{blue}{update\_pangle2\_classicmerge.sh}}
		\label{Pipeline_Functions_pangle}

{\bf update\_pangle2\_classicmerge.sh} is a small variation on the script {\bf update\_pangle2.sh}, which was written by Sean McManus to compile the derived GONG-network image angles into a format familiar to the GRASP routine CAMOFFSET (see Appendix~\ref{App_CAMOFFSET} for basic documentation), and then to call CAMOFFSET to write the site/time-appropriate OFFSET-keyword values into the headers of all FITS files in an indicated directory.  The two primary differences between the GONG ClassicMerge call to this script and the GONG+ pipeline call are that:
\begin{list}{}{}
\item {\bf a)} The script is called with the -c flag to indicate that this is GONG-Classic data and, therefore, to direct CAMOFFSET not to include Ronchi corrections when computing the OFFSET angles.
\item {\bf b)} This ClassicMerge version only calls CAMOFFSET if the input FITS image does {\em not} have the FILLED keyword in its header.  If, instead, the FILLED keyword is present, {\bf update\_pangle2\_classicmerge.sh} deletes the file from the GONG ClassicMerge working directory.  (This is done here in order to help streamline the early-stage processing.)
\end{list}

The OFFSET keyword in a GONG header indicates the computed residual misalignment (in degrees) between solar-north and the y-axis of an image.  This keyword is used by IMMERGE to correct the rotation of input site images before merging them (therefore, by definition, merged images have OFFSET = 0.0).

While this script is set up to update the OFFSET values of all FITS files in a single directory (covering one UT-day for one site), CAMOFFSET does seek to provide appropriate OFFSET values based on each file's {\em site-local} date.  Therefore, in calling {\bf update\_pangle2\_classicmerge.sh} in the GONG ClassicMerge pipeline, it is necessary to ensure that the {\bf coy} (computed angles coefficients) files for days adjacent (depending on the site, as outlined in \S\ref{Pipeline_Functions_sdlist}) to the of-interest UT date have also been pulled from the GONG-Classic archive and are available to the GONG ClassicMerge working directory.

		\subsubsection[\textcolor{blue}{Miscellaneous Small Functions}]{\textcolor{blue}{Miscellaneous Small Functions}}
		\label{Pipeline_Functions_Misc}

\noindent {\bf \textcolor{black}{gong\_classicmerge\_headeradd.cl}:  }
This script is essentially just a copy of a code snippet sourced from {\bf velfit.cl}, allowing the user to specify one header keyword, one value for that keyword, and a list of FITS files to which that header keyword and value should be added.  In the GONG ClassicMerge pipeline, it is used to add the ELEV (site elevation) keyword to the image headers.

$\,$\\
\noindent {\bf \textcolor{black}{gong\_classicmerge\_ha\_scalecheck.cl}:  }
This is a small script that {\bf a)} takes in a list of FITS files, {\bf b)} reads the value of the HA (hour angle) keyword in each file header, and {\bf c)} rewrites that value as HA$_\mathrm{out}$ = HA$_\mathrm{in}$ - $2\pi$ in the cases where HA$_\mathrm{in} > \pi$.  It is run on all of the GONG-Classic images after they have been pulled from the archive in order to ensure that {\bf velfit.cl} can easily sort/exclude observations according to their HA values.

$\,$\\
\noindent {\bf \textcolor{black}{get\_parm}:  }
This is a small command-line function available within the GONG pipelines via:\\
\indent source {\bf std\_bash\_lib.sh}.\\
Given the name of a FITS file (which may be compressed) and a header keyword, it will return the value of that keyword.


$\,$\\
\noindent {\bf \textcolor{blue}{FITSWASH}:   }
This is a function used within the GONG pipelines to update and tidy processed images' FITS headers, which, in the GONG ClassicMerge pipeline, gets applied to the {\em merged} images only.  It requires an input FITS filename plus a FITS-header template file.  For the GONG ClassicMerge pipeline, the template used is called ``gong\_classicmerge\_velmerge\_header.kw" and primarily differs from the standard GONG+ merged-magnetogram template in that it specifies these merged GONG-Classic data are {\em velocity} images.


\section[\textcolor{blue}{Data Checks and Exceptions}]{\textcolor{blue}{Data Checks and Exceptions}}
\label{DataChecks}

	\subsection[\textcolor{blue}{Basic Checks}]{\textcolor{blue}{Basic Checks}}
	\label{DataChecks_Basic}

During the operation of the GONG ClassicMerge pipeline, the code performs various checks to determine which data to include in the final merged product.  These checks include:
\begin{enumerate}
   \item The first and most obvious check is one designed to gauge whether a given file has been corrupted or is otherwise unreadable.  If an image file copied to the working directory could not be gunzipped, or if the NAXIS1 keyword could not be read out of the image header, then the observation is discarded.  Out of the full six-years worth of GONG-Classic data, such discards were necessary on eight UT-dates:
      \begin{list}{-}{}
         \item One corrupted observation was removed from each of the following data directories: ctvzi961217, mlvzi970718, tdvzi981026, bbvzi000105, ctvzi001122, and levzi000830.
         \item 464 corrupted observations were removed from tdvzi981027.
         \item 472 corrupted observations were removed from tdvzi981028.
      \end{list}
   \item As the VELSCALE keyword is necessary for calibrating the site data before merging, and the {\bf qac} files are necessary for computing VELSCALE values, the next check made is to ensure that the {\bf qac} file is available for each site-day worth of observations.  If for some reason a given {\bf qac} file is not available, then all observations taken on that site day are removed from the GONG-Classic merging.  An exception to this rule was made during the data processing for a swath of observations taken in the early fall of 1999, and is discussed in \S\ref{DataChecks_QAC}.  Site days removed from GONG ClassicMerge processing due to missing {\bf qac} files include:
      \begin{list}{-}{}
      \item 3 site days at Big Bear:  bbqacV990806--990808
      \item 5 site days at Cerro Tololo: ctqacV990801--990805
      \item 14 site days at Learmonth: leqacV990803--990812,  990814--990817
      \item 6 site days at Mauna Loa: mlqacV960910, 980328, 980424, 990801, 990803, 990805
      \item 13 site days at Udaipur: udqacV970412--970419, 970421--970425
      \end{list}
   \item As discussed in \S\ref{Pipeline_Functions_reject}, the {\bf gong\_classicmerge\_reject} code was run on all of the input observations to check that basic standards in GONG telescope pointing were being met.  As a result of those checks: 
      \begin{list}{-}{}
         \item 1325 observations were excluded from the 1995 GONG-Classic merging (886 CT, 5 ML, 12 TD, and 422 UD observations).
         \item 3016 observations were excluded from 1996 (13 CT, 3 LE, 1 ML, 59 TD, and 2940 UD).
         \item 1137 observations were excluded from 1997 (1115 CT, 1 LE, 3 ML, and 18 UD).
         \item 2 observations were excluded from 1998 (1 LE, and 1 TD).
         \item 539 observations were excluded from 1999 (46 BB, 227 CT, 190 LE, 72 TD, and 4 UD).
         \item 26,377 observations were excluded from 2000 (25,266 BB, 473 LE, and 638 ML).
         \item 5955 observations were excluded from 2001 (all BB observations).
      \end{list}
\end{enumerate}

	\subsection[\textcolor{blue}{Instances of Problematic VELSCALE}]{\textcolor{blue}{Instances of Problematic VELSCALE}}
	\label{DataChecks_VELSCALE}

Once the GONG ClassicMerge pipeline was run, a number of instances were discovered where the computed VELSCALE value was a clear outlier from the norm (in general, for GONG-Classic data, VELSCALE should be quite close to 1.0 $\pm$ 0.1).  Inspection of some example images found no obvious cause for these discrepancies; however, the outlier VELSCALE values were {\bf not} the correct values needed to calibrate those observations.  Therefore, a second merge processing was performed on the 50 affected days, in which:
   \begin{enumerate}
      \item The site-day observations containing incorrect VELSCALE values in their headers were removed from the working directories.
      \item IMMERGE was re-run on the remaining observations.
      \item {\bf get\_parm} was re-run on the new mrvzi files to output a new GCM\_parm\_mr\_NSITES.dat file, needed to compute the revised duty cycle.
   \end{enumerate}
In all, this re-processing was performed on:
   \begin{list}{-}{}
      \item 2 UT days in 1995 (both due to UD VELSCALE values).
      \item 7 UT days in 1996 (3 due to CT, 1 due to LE, and 3 due to TD values).
      \item 7 UT days in 1997 (1 due to CT, 2 due to LE, 2 due to ML (single site day), 1 due to TD, and 1 due to UD values).
      \item 4 UT days in 1998 (1 due to LE, 2 due to TD, and 1 due to UD values).
      \item 27 UT days in 2000 (25 due to BB, 1 due to CT, and 1 due to ML values).
      \item 3 UT days in 2001 (2 due to BB and 1 due to LE values).
   \end{list}

A large number of the cases where VELSCALE was computed incorrectly for a given site day reported very few observations available for that site day.  Therefore, once the re-merge was completed, the duty cycle was found to have dropped by less than 2\% in the majority of cases.  The exceptions noted are:
   \begin{list}{-}{}
      \item 951011: The duty cycle fell from 0.93 to 0.87 (6\%).
      \item 960706: The duty cycle fell from 0.97 to 0.93 (4\%).
      \item 960922: The duty cycle fell from 0.55 to 0.45 (10\%).
      \item 970108: The duty cycle fell from 0.87 to 0.84 (3\%).
      \item 980807: The duty cycle fell from 0.71 to 0.68 (3\%).
      \item 000103: The duty cycle fell from 0.14 to 0.06 (8\%).
      \item 000810: The duty cycle fell from 0.96 to 0.94 (2\%).
   \end{list}

	\subsection[\textcolor{blue}{Missing Angles from 1999}]{\textcolor{blue}{Missing Angles from 1999}}
	\label{DataChecks_COY}

The duty cycle of the GONG ClassicMerge data (see Figure~\ref{FIG_BP_dutycycle}) drops to 0.0 for the period of 1999-05-30 through 1999-08-01.  This is due to a complete lack of {\bf coy}-angle files (necessary for computing the correct angular orientation of each observation) within this time-period.  This lack is almost certainly due to a corresponding lack of driftscans calibration data \citep{Toner2001} taken during this time.  However, until such time as new {\bf coy}-angles can be extrapolated, the GONG-Classic observations taken in June and July of 1999 cannot be merged.  (As a side note, the {\bf qac} files are also missing during this period, and continue so into October of 1999, as discussed in \S\ref{DataChecks_QAC}.)

	\subsection[\textcolor{blue}{Missing {\bf qac} from 1999}]{\textcolor{blue}{Missing {\bf qac} from 1999}}
	\label{DataChecks_QAC}

In 1999, there are nearly three months (beginning of August through most of October) during which {\bf coy}-angle results have returned (see \S\ref{DataChecks_COY}), but there are still no {\bf qac} files stored in the archive, with the exception of those for the site at Teide.  However, beginning in early August, the archived site data begin to have VELSCALE values already written into their headers.  Therefore, for the data processed from August through October of 1999, site-day observations that are missing a corresponding {\bf qac} file are {\em not} discarded {\em if} they already have VELSCALE in their headers.

This means that, if there has been a change in the {\bf velfit} algorithm between when those VELSCALE values were written and when the GONG ClassicMerge processing was done in 2015, then, depending on the date and site, the VELSCALE value used for image callibration may have been computed one of two ways.  (This potential dichotomy is probably similar to the site-to-site variation in the VELSCALE computation that also happened at the beginning of 1996, as discussed in \S\ref{Pipeline_Functions_velfit}.)  The epochs for VELSCALE in the GONG ClassicMerge data are as follows:
   \begin{list}{-}{}
      \item Big Bear: Archived, in-header VELSCALE beginning 1999-08-06; returned to {\bf qac}-velfit VELSCALE 1999-10-18.
      \item Cerro Tololo: Archived VELSCALE from 1999-08-02 to 1999-10-26.
      \item Learmonth: Archived VELSCALE from 1999-08-02 to  1999-10-24.
      \item Mauna Loa: Archived VELSCALE from 1999-08-03 to 1999-10-13.
      \item Teide: Standard {\bf qac} VELSCALE throughout.
      \item Udaipur: Standard {\bf qac} VELSCALE throughout (due to monsoon, there are no observations in this period until 1999-10-06).
   \end{list}

\section[Post-Processing Quality Checks]{Post-Processing Quality Checks}
\label{PPAnalysis}

All merged dopplergrams were additionally subjected to the following post-processing quailty checks:
%
%

$\,$

\begin{figure}[t]
     \begin{center}
     \includegraphics[scale=0.7]{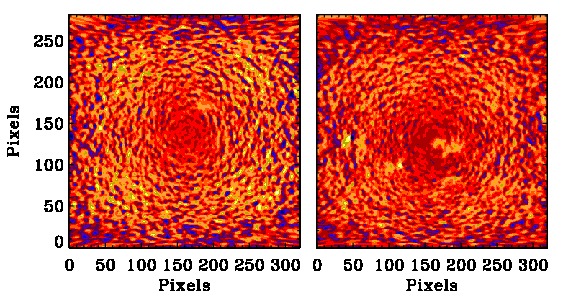}
     \caption[Example Mean-images Quality Check]{Example Mean images used in our initial quality checks: 1995-06-01 (low activity; left) \&  2001-01-24 (high activity; right).}
     \label{FIG_Kiran1}
     \end{center}
     \end{figure}

\noindent {\bf 1. Visual inspection of Mean and Powermap images: }  Movies of Mean and Powermap images made for each day of observations were visually inspected.
For these images, the central regions of the merged images ($\mathrm{lat}_\mathrm{bound}=\pm78^\circ$; $\mathrm{long}_\mathrm{bound}=\pm56^\circ$) were remapped onto a sin(lat)-longitude grid, with tracking performed at the surface differential rotation rate relative to the noon time for each day (i.e., by selecting the image at 12:00:00 UT as the reference).
These tracked cubes were used to generate daily image means and power maps, where:  
\begin{Verbatim}[commandchars=\\\{\}]
                 Mean = SUM\textsubscript{fullday}[ tracked.image ] / N\textsubscript{images},
             Powermap = SUM\textsubscript{fullday}[ (tracked.image - Mean)\textsuperscript{2} ] / N\textsubscript{images}.
\end{Verbatim}
Examples of Mean images at periods of low and high solar activity are shown in Figure~\ref{FIG_Kiran1}.

Careful visual inspection of the progression of these Mean and Powermap images allowed for identification of outlier pixels and/or errant spatial trends.
After identifying a bad day, each corresponding input image was inspected separately, and bad images were replaced with a blank image in the final ClassicMerge product.  
Note that this technique was also used during the ClassicMerge code-development process to identify and correct early errors in our image-rejection algorithms (described in \S\ref{DataChecks_Basic}).

$\,$

\noindent {\bf 2. Inspection of daily $\ell$-$\nu$ diagrams: }  Daily $\ell$-$\nu$ diagrams were also inspected for anomalies.  These diagrams were generated using the spherical harmonic time series only for those days when the duty cycle was $> 65\%$.

\begin{figure}
     \begin{center}
     \includegraphics[scale=0.65]{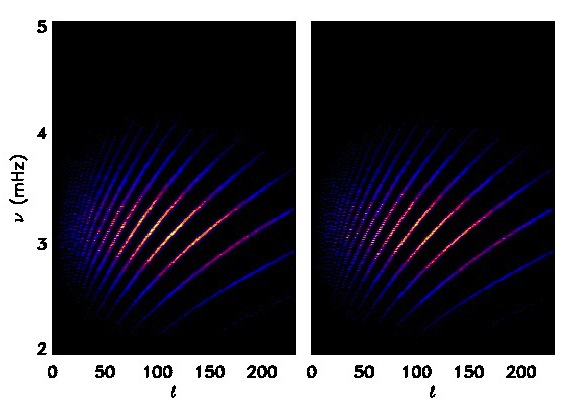}
     \caption[Example $\ell$-$\nu$-diagram Quality Check]{Example $\ell$-$\nu$ diagrams used in our second-pass quality checks: 1995-06-01 (left) \&  2001-01-24 (right).}
     \label{FIG_Kiran2}
     \end{center}
     \end{figure}

For this test, the first step was to perform spherical-harmonic decomposition on each merged image, using a setting of $\ell_\mathrm{max}=240$.  
Next, for each UT day inspected, the time series of spherical harmonic coefficients was Fourier transformed to produce $m$-$\nu$ power spectra.  These $m$-$\nu$ spectra were corrected for solar rotation and averaged over $m$ to produce an $\ell$-$\nu$ diagram for that day.
Two examples of $\ell$-$\nu$ diagrams thus generated are shown in Figure~\ref{FIG_Kiran2}.
Finally, all of the generated $\ell$-$\nu$ diagrams were stacked into a master time-series spanning the duration of GONG ClassicMerge observations, which was then visually inspected in cross-section at several $\ell$ values.  

\begin{figure}
     \begin{center}
     \includegraphics[scale=0.19]{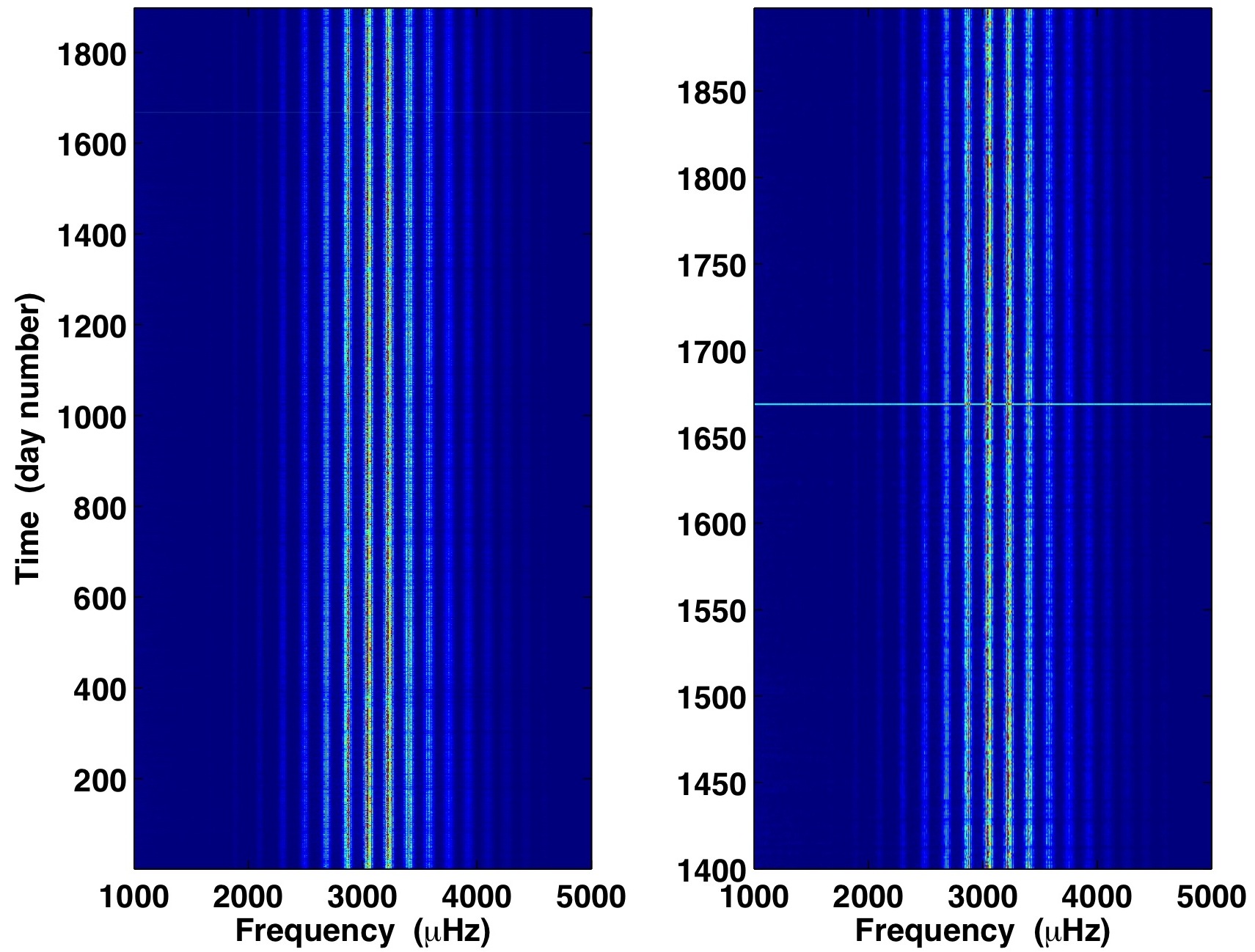}
     \caption[Example $\ell$-$\nu$-time-series Quality Check]{Stack plots of $\ell$-$\nu$ diagrams for $\ell=50$.  A faint horizontal line is clearly visible at the time sequence=1665.  The left image is for the entire period of GONG-Classic data, and the right image covers the period around the horizontal line.}
     \label{FIG_Kiran3}
     \end{center}
     \end{figure}

Only one bad image was found and discarded as a result of the $\ell$-$\nu$ inspection test, which can be seen in the time-series map for $\ell=50$ in Figure~\ref{FIG_Kiran3}.
A closer look reveals a horizontal line at about time sequence = 1665.  This implies that there was high background noise present in the $\ell$-$\nu$ diagram corresponding to that particular day.  
Follow-up inspection of that day's individual input images yielded the dopplergram responsible.
This image was replaced in the final ClassicMerge product with a blank image and given the header keyword FILLED, which is the standard GONG-pipeline flag for image exclusion.

$\,$

\noindent Based on these tests, the GONG project is releasing the ClassicMerge data product to the user community (http://nisp.nso.edu/data).

\section*{Acknowledgement}

The authors would like to thank Manuel Diaz Alfaro (IAC, Spain) for his feedback and help with early analysis of the quality of the ClassicMerge data.

\begin{list}{}{}
\item $\,$
\item $\,$
\end{list}
\appendix
\section{Appendix}
\subsection{CAMMOFFSET help page}
\label{App_CAMOFFSET}

Below is a transcription of the text that appears at: http://gong.nso.edu/data/iraf\_help/camoffset.html as accessed on December 31st, 2015 \citep{citeCAMOFFSET}.

\begin{verbatim}

CAMOFFSET (Aug96)            	pipeline            CAMOFFSET (Aug96)



NAME
    camoffset  --  Calculate  camera  offset  angles  using  COPIPE   or 
    DRIFTSCAN results.
    
    
USAGE
    camoffset input
    
    
PARAMETERS
    
    input
        List  of  input  V,  M,  or I images for which the camera offset
        angle is to be determined.
    
    drift_scan = no
        Use Drift-Scan formuli only.
    
    tables = "grasplib$SITES/"
        Path to offset tables.
    
    verbose = no
        Verbose output to STDOUT.
    
    
DESCRIPTION
    
    camoffset uses the hour angle from the input image headers
    (keyword:  HA;  units:  radians) and empirical formula to determine
    the camera offset angle.  The calculated offset angle in degrees is
    written into the image header as the value of the keyword OFFSET.
    
    The GONG pipeline process COPIPE determines an internally consistent
    set of equations (maximum 6  term  power  law)  describing  how  the
    camera   offset   angle   changes  with  time  at  each  site.   The 
    coefficients for these equations are stored in STSDAS binary  tables
    located in tables.  The table columns are:
    
         gps_date         CH*10        \%10s "GPS Date"      
         n_coeff          I             \%1d "No. of Coefficients"
         c1               R              \%f ""              
         c2               R              \%f ""              
         c3               R              \%f ""              
         c4               R              \%f ""              
         c5               R              \%f ""              
         c6               R              \%f ""              
         drift_scan       CH*2          \%2s "PM Driftscan"  
    
    The  column  "drift_scan"  is  blank  except for those dates where a
    drift scan was taken, in which case the column contains "pm".
    
    cam offset uses the value of the header keyword SITE to determine
    ensure that the proper table is use.  Accepted values of SITE are
    
         tc ==> Tucson 
         bb ==> Big Bear
         ml ==> Mauna Loa
         le ==> Leamonth
         ud ==> Udaipur
         td ==> Teide
         ct ==> CTIO
    
    If drift_scan=yes, the coefficients used are those from the most
    recent drift scan (drift_scan == "pm") taken prior to  the  date  in
    question.
    
    If verbose=yes, the calculated offset angles are printed to STDOUT
    as the task executes.
    
    
EXAMPLES
    
    1.  Determine the camera offset angles for a set of Teide calibrated
        velocity images.
    
        pi> camoffset tdvci*.imh
    
    
    2.   Deterine the camera offset angles for a set of Teide calibrated
        velocity images using only the drift scan results.
    
        pi> camoffset tdvci*.imh drift_scan+
    
    
TIME REQUIREMENTS
    
    Nothing significant.
    
    
BUGS
    
    
    
SEE ALSO
    
    imoffset, xoffset

\end{verbatim}

\begin{list}{}{}
\item $\,$
\item $\,$
\item $\,$
\item $\,$
\end{list}
\subsection{EPHEMINTERP help page}
\label{App_EPHEMINTERP}

Below is a transcription of the text that appears at:  http://gong.nso.edu/data/iraf\_help/epheminterp.html as accessed on December 14th, 2015 \citep{citeEPHEMINTERP}.

\begin{verbatim}

EPHEMINTERP (May96)            ephem           EPHEMINTERP (May96)



NAME
    epheminterp  --  Interpolate  ephemeris  tables  to the date/time of
    observation.
    
    
USAGE
    epheminterp input
    
    
PARAMETERS
    
    input
        Input image list.
    
    ephem_dir = "ephem$"
        Path  to  the  directory   containing   the   ephemeris   tables 
        subdirectories.
    
    update_hdr = yes
        Update  the  image  header  to include B0, L0, PANGLE, SEMIDIAM,
        VCOR[1-4].
    
    refraction = no
        Include  atmospheric  refraction  in  the  velocity   correction 
        calculation.
    
    verbose = no
        Output results to STDOUT.
    
    
DESCRIPTION
    
    EPHEMINTERP  uses  cubic  interpolation  to  determine the following
    quantities for each input image:
        L0, B0    - Heliographic coordinates of sub-earth point (degrees),
        P_ANGLE   - Solar position angle (degrees),
        SEMIDIAM  - Solar Semi-diameter (degrees),
        VCOR[1-4] - Observer's velocity correction (m/s) coefficients. 

    If update_hdr = yes, these quantities are written into the image
    header  using  keywords  B0,  L0,  P_ANGLE,  SEMIDIAM, VCOR1, VCOR2,
    VCOR3 and VCOR4.
    
    If verbose = yes, these results are also written to STDOUT.
    
    If refraction = yes, atmospheric refraction is taken into account    
    when  the  observer's velocity correction is determined.  The zenith
    distance (z = 90 - a, a = altitude) decreases by an amount  R  given
    by:
    
        R (degrees) = 0.00452 * P / ((273+T)*tan(a))        a > 15 degrees
    
                   P*(0.1594 + 0.0196*a + 0.00002 * a**2)
        R (deg) = ----------------------------------------  a <= 15 degrees
                  [(273 + T)(1 + 0.505*a + 0.0845 * a**2)]
    
    
        where  P == Pressure in millibars
               T == Temperature in degrees Celcius
               a == altitude in degrees
    

    The ephemeris tables are located in subdirectories of ephem_dir
    called jYYYY where YYYY is the year of observation.  Each
    subdirectory contains three ephemeris files:
    
        sunapp.dat - apparent geocentric postion and velocity of the sun.
        sunpbl.dat - heliographic coordinates, p-angle and semi-diameter.
        utgmst.dat - UT to GMST time conversion.
    
    These  tables range from December 31 of the previous year to January
    2 of the following year which allows cubic intepolation on Janary  1
    and December 31 of the year of observation.
    
    EPHEMINTERP requires the following header keywords be in the input
    images:
    
       KEYWORD       FORMAT              Description
       =====================================================================
       DATE-OBS  YYYY/MM/DD    GPS Date of observation.
       TIME-OBS  HH:MM:SS.SSS  GPS Time of observation.
       LAT       double        Latitude of observer (radians north)
       LON       double        Longitude of observer (radians east)
       ELEV      double        Elevation of observer (meters)(default = 0.)
       BAROMET   real          Barometric pressure (millibars)(default = 0.)
       TEMP_OUT  real          Temperature (degrees Celsius)(default = 0.)
    
    Failure to access one of the  above  keywords  will  result  in  the
    default  value being used, or if there is no default, termination of
    the EPHEMINTERP process.
    
    
    The values of B0, LO, P_ANGLE and SEMIDIAM are  obtained  by  direct
    cubic   interpolation  in  the  "sunpbl"  table,  which  is  indexed 
    according to TDT (Terrestrial Dynamical Time)  since  the  beginning
    of  the  year.   TDT  is obtained from the GPS date and time via the
    following relation:
    
        TDT = GPS + 51.184 seconds
    
    
    The values of VCOR[1-4] are obtained by using an  ephemeris  program
    which   requires   as  input  the  site  latitude  (radians  north), 
    longitude (radians east) elevation  (meters),  temperature  (degrees
    C),  pressure  (millibars),  P-angle (degrees), geocentric positions
    (au),  geocentric velocities (au/day),  and  the  Greewich  Apparent
    Sidereal Time (hours).
    
    NOTE:  The position angle is set to 0. for the velocity correction
           calculation.  Since the GONG instrument is attempting to 
           compensate for the P_ANGLE, the camera offset angle will need 
           to be used to properly remove the velocity gradients.
    
           To remove the observer's velocity at each pixel, use the formula
           
                 vcor1 + vcor2*X + vcor3*Y + vcor4*(X**2 + Y**2)/2.
    
           for  -1 <= X <= +1;  ABS(X)=1 at the limb
                -1 <= Y <= +1;  ABS(Y)=1 at the limb
    
           This formula assumed that the Y-axis is aligned with Heliographic
           north.  If there is a rotation about the line site away from this
           assumption then VCOR2 and VCOR3 must be adjusted accordingly:
    
                    crot = cos(rotation_angle)
                    srot = sin(rotation_angle)
                    tmp2  = vcor2*crot + vcor3*srot
                    tmp3  =-vcor2*srot + vcor3*crot
                    vcor2 = tmp2
                    vcor3 = tmp3
    
    
    The  geocentric  positions  and velocities of the sun are determined
    by interpolation the "sunapp" table  which  is  to  TDT.   The  GAST
    (Greenwich  Apparent  Sidereal  Time)  is obtained by looking up the
    the GMST (Greenwich Mean Sidereal Time)  for  0h  and  applying  the
    following formula:
    
        gast = gmst0 + utc_hour + eqnxtrm/3600.0d0
    
        where gmst0     == Greenwich Mean Sidereal Time at 0h UT
              utc_hour  == Number of UT hours since 0h UT * 1.0027379093d0
              eqnxstrm  == Equation of equinoxes interpolated at utc_hour.
    
    where  the constant (1.0027379093d0) converts from solar to siderial
    time.
    
    The task uses  UTC  (Coordinated  Universal  Time)  rather  than  UT
    (maximum   difference   is   1  second)  because  it  is  easier  to 
    determine.  UTC may have a leap second introduced at the end of  the
    months of December and June, depending on the evolution of UT1-TAI.
    
        UTC = GPS + 19s - delta
    
    where  "delta"  is  an  integral  second  offset  determined by NEOS
    (National Earth Orientation Service), which has  an  email  bulletin
    (Bulletin  C)  sent  out  every  6  months  to  confirm  or deny the
    addition of a leap second.
    
    
    
EXAMPLES
    
    1)  Run EPHEMINTERP on  all  the  Mauna  Loa  data  in  the  current
    directory.
    
        ep> epheminterp ml*.imh
    
    
    
TIME REQUIREMENTS
    
    It  takes  approximately 15 seconds clock time to do 680 images on a
    SPARCstation 20.
    
    
    
BUGS
    
    None known.
    
    
    
REFERENCES
    
    Astronmical Almanac, Section B (Time Scales).
    
    
SEE ALSO
    ephvel, ephgeo

\end{verbatim}

\begin{list}{}{}
\item $\,$
\item $\,$
\item $\,$
\item $\,$
\end{list}
\subsection{FNDLMB help page}
\label{App_FNDLMB}

Below is a transcription of the text that appears at:  http://gong.nso.edu/data/iraf\_help/fndlmb.html as accessed on December 31th, 2015 \citep{citeFNDLMB}.

\begin{verbatim}

FNDLMB (Sep94)                   gongcor                  FNDLMB (Sep94)



NAME
    fndlmb - Determine the figure of the limb of the solar image
    
    
USAGE
    fndlmb input output usehdr updhdr
            xcenter ycenter minorax majorax angle
            nfit width thresh pcfit pixlenx pixleny prntsw nfndlmb
    
    
PARAMETERS
    
    input
        List of input pixel arrays containing full disk  images  of  the
        sun.
        If N1 and N2 are the dimensions of the images,
        then N1/4.LE.N2.LE.4*N1; i.e., the dimensions must
        be approximately equal.
    
    output
        List  of  second  derivative  (Laplacian)  images. The number of
        output images may be zero or must be the same as the  number  of
        input images.
        The  output  image  names  may  not  be equal to the input image
        names.
    
    usehdr
        If YES, the initial estimates for  the  ellipse  parameters  are
        taken  from  the  values  found  in the input image header.  The
        values of the five ellipse input parameters described below
                XCENTER YCENTER MINORAX MAJORAX ANGLE
        are ignored.
    
    updhdr
        If YES, the values of the ellipse parameters in the header  will
        be updated after the parameters are determined.
    
    xcenter 
        Estimate  for  the  x  co-ordinate  of the center of the ellipse
        describing the solar image.   The  x  axis  corresponds  to  the
        first dimension of the pixel array.
        N1 is the first dimension of the image.
        Nominal value might be 0.5*N1.
        Units are pixels.
        1.LE.XCENTER.LE.N1
    
    ycenter 
        Estimate  for  the  y  co-ordinate  of the center of the ellipse
        describing the solar image.   The  y  axis  corresponds  to  the
        second dimension of the pixel array.
        N2 is the second dimension of the image.
        Nominal value might be 0.5*N2.
        Units are pixels.
        1.LE.YCENTER.LE.N2
    
    minorax
        Estimate  for  the  semi-minor axis of the ellipse descibing the
        solar image.  N1,N2 are the dimensions of the image.
        Nominal value might be 0.4*MIN(N1,N2).
        Units are pixels.
        2.LE.MINORAX.LE.N1-1
    
    majorax
        Estimate for the semi-major axis of the  ellipse  descibing  the
        solar image.  N1,N2 are the dimensions of the image.
        Nominal value might be 0.4*MIN(N1,N2).
        Units are pixels.
        2.LE.MAJORAX.LE.N2-1
    
    angle
        Estimate  for  the  ellipse rotation angle.  A counter-clockwise
        rotation of the y axis to the ellipse major axis is  a  positive
        angle.
        Units are degrees.
        -90.0.LE.ANGLE.LE.90.0
    
    nfit
        Number of parameters determined by the least squares fit.
        If  NFIT=4;  XCENTER,  YCENTER,  MINORAX, MAJORAX are determined
        for an ellipse with rotation angle, ANGLE, the input parameter.
        If NFIT=5; XCENTER, YCENTER, MINORAX,  MAJORAX,  and  ANGLE  are
        determined for the ellipse.
        NFIT=4,5
    
    width
        Width  of  the  band  of second derivatives searched by the zero
        crossing algorithm.
        If N1,N2 are the dimensions of the image.
        A nominal value might be 0.03*MIN(N1,N2).
        Units are pixels.
        2.LE.WIDTH.LE.MIN(MINORAX,MAJORAX)-1
    
    thresh
        Threshold of the band of  second  derivatives  searched  by  the
        zero  crossing  algorithm.  A reasonable value may be determined
        by inspection or statistical analysis of the  second  derivative
        image.
        A nominal value for 16-bit intensity images might be 1000.
        If THRESH=0; the value of THRESH will be automatically
        determined to be one-half of the maximum value of the
        second derivatives of the center row and center column
        of the image.
        0.0.LE.THRESH
    
    pcfit
        Second  derivative  zero  crossings  within PCFIT percent of the
        ellipse determined by the first fit are retained for the  second
        fit.  This parameter is used to reject spurious points.
        A nominal value would be 1.0; i.e., 1%.
        0.1.LE.PCFIT.LE.10.0
    
    pixlenx pixleny
        Pixel length in x-direction and y-direction in arbitrary units.
        Used  if  USEHDR=NO,  otherwise  the task attempts to read these
        parameters from the image headers.
        If UPDHDR=YES, the task will  update  these  parameters  in  the
        image headers.
    
    prntsw
        Switch  that  controls  writing the locations (in pixels) of the
        zero crossings to the standard output.
        If NO, zero crossings are not output.
    
    nfndlmb
        The number of times the FNDLMB algorithm will be  invoked.   The
        limb  parameters (XCENTER, YCENTER, MINORAX, MAJORAX, ANGLE) and
        the THRESH that were determined during  the  previous  call  are
        used  as  the  estimates  for  the  next call.  NFIT, WIDTH, and
        PCFIT as specified by the input  parameters  are  used  for  all
        invocations of the algorithm.
    
    
IMAGE HEADER WORDS
    
    IF  USEHDR  = YES, the following header parameters are read from the
    input image header:
        PIXLENX, the x-direction size of the camera pixels
        PIXLENY, the y-direction size of the camera pixels
        FNDLMBXC, the x-coordinate of the center of the disk in pixels
        FNDLMBYC, the y-coordinate of the center of the disk in pixels
        FNDLMBMI, the length of the semiminor axis in pixels
        FNDLMBMA, the length of the semimajor axis in pixels
        FNDLMBAN, the ellipse rotation angle in degrees
    
    IF UPDHDR = YES, the following header parameters are written to  the
    input image header:
        PIXLENX = PIXLENX, an input parameter
        PIXLENY = PIXLENY, an input parameter
        FNDLMBXC, the x-coordinate of the center of the disk in pixels
        FNDLMBYC, the y-coordinate of the center of the disk in pixels
        FNDLMBMI, the length of the semiminor axis in pixels
        FNDLMBMA, the length of the semimajor axis in pixels
        FNDLMBAN, the ellipse rotation angle in degrees
    
    IF  UPDHDR = YES, the task writes the processing parameters into the
    input header using 'gr_history' from 'grutil/'.
    
    IF UPDHDR = YES and the fitting algorithm fails, FILLED   =  NO,  is
    added  to  the image header and the limb parameters are set to their
    input values, the initial estimates.
    
    
DESCRIPTION
    
    FNDLMB uses a modified version of  Stuart  Jefferies'  limb  finding
    algorithm  (with  the  addition  of  the  ellipse rotation angle) to
    obtain an ellipse  which  best  approximates  (in  a  least  squares
    sense) the limb of the solar image.
    
    In  April,  1992, FNDLMB was modified to include features from Jesus
    Patron's  version  of  the  limb  finder.   These  changes  included 
    searching  for  zero-crossings  in  both  the  x  and  y  directions 
    (previously, the search was in the x direction); searching from  the
    outside  of the solar disk to the inside (previously, the search was
    through increasing x coordinates); and some  details  of  the  logic
    associated with detecting an acceptable zero-crossing were changed.
    
    The  ellipse (as described by the center co-ordinates and semi-minor
    and semi-major axes and rotation angle) is written to  the  standard
    output and to the  image header.
    
    If  the  input image consists of more than 1 2d array of pixels, the
    2d arrays are summed into 1 2d array before the  ellipse  parameters
    are determined.
    
    Assume that the solar limb can be fit by an ellipse defined by
            XC - center x-coordinate in pixels, first pixel dimension.
            YC - center y-coordinate in pixels, second pixel dimension.
            AE - minor ellipse width along the rotated x-axis in pixels.
            BE - major ellipse width along the rotated y-axis in pixels.
            ANG - rotation angle in degrees
                    clockwise rotation of the y-axis is a negative ANG.
    
    
                                 .
                                 .- (-ANG) - 
                                 .       BE
                                 .      BE
                                 .     BE  
                                 .    BE     
                                 .   BE       
                                 .  BE        
                         AE      . BE        
                             AE  .BE        
       . . . . . . . . . . . .(XC,YC). . . . . . . . . . . . . X
                                BE  AE    
                               BE.      AE
                              BE .          
                             BE  .   
                            BE   . 
                           BE    .
                          BE     .
                         BE      .
                                 .
                                 Y
    
    Definitions:
            N1,N2 - pixel array dimensions
            DX=X-XC
            DY=Y-YC
            COSA=COS(ANG*PI/180.0)
            SINA=SIN(ANG*PI/180.0)
            CXX=((COSA/AE)**2+(SINA/BE)**2)
            CXY=-2*COSA*SINA*(1.0/AE**2-1.0/BE**2)
            CYY=((SINA/AE)**2+(COSA/BE**2)
    
    Then the ellipse has the form
            CXX*DX**2-CXY*DX*DY+CYY*DY**2=1.0
    
    An initial estimate for the ellipse is provided by 
    the input parameters:
            XC,YC,AE,BE,ANG.
    
    Determine a search band(two concentric ellipses) centered on 
    the ellipse defined by the input parameters and the 
    search band width, WIDTH:
    
      The search band is defined for both the x and y coordinates.  
      The band in terms of the x coordinates is discussed first.
    
      For every Y index from 2 to N2-1 find 
        XOL-outside of search band, lower x value
        XOH-outside of search band, higher x value
        XIL-inside of search band, lower x value
        XIH-inside of search band, higher x value
    
            For XOL and XOH, use
                    AEO=AE+WIDTH
                    BEO=BE+WIDTH
                    to calculate CXXO,CXYO,CYYO.
            For XIL and XIH, use
                    AEI=AE-WIDTH
                    BEI=BE-WIDTH
                    to calculate CXXI,CXYI,CYYI.
    
            ARGO =CXYO**2*DY**2-4*CXXO*(CYYO*DY**2-1)
            ARGI =CXYI**2*DY**2-4*CXXI*(CYYI*DY**2-1)
            IF(ARGO.LT.0)XOH & XOL are undefined
            IF(ARGI.LT.0)XIH & XIL are undefined
    
            DXOL=(CXYO*DY-SQRT(ARGO))/(2*CXXO)
            DXOH=(CXYO*DY+SQRT(ARGO))/(2*CXXO)
            XOL=DXOL+XC
            XOH=DXOH+XC
            XOL=FLOOR  (XOL)
            XOH=CEILING(XOH)
            DXIL=(CXYI*DY-SQRT(ARGI))/(2*CXXI)
            DXIH=(CXYI*DY+SQRT(ARGI))/(2*CXXI)
            XIL=DXIL+XC
            XIH=DXIH+XC
            XIL=CEILING(XIL)
            XIH=FLOOR  (XIH)
    
            One of the x values defining the inside or outside of the
            search band must be in the pixel array.
                    IF(XOL.GT.N1-1)XOH & XOL are undefined
                    IF(XIL.GT.N1-1)XIH & XIL are undefined
                    IF(XOH.LT.   2)XOH & XOL are undefined
                    IF(XIH.LT.   2)XIH & XIL are undefined
    
            If one is in the pixel array, make sure the other is 
            as well.
                    IF(XOL.LT.   2)XOL=   2
                    IF(XIL.LT.   2)XIL=   2
                    IF(XOH.GT.N1-1)XOH=N1-1
                    IF(XIH.GT.N1-1)XIH=N1-1
    
      For every X index from 2 to N1-1 find 
        YOL-outside of search band, lower y value
        YOH-outside of search band, higher y value
        YIL-inside of search band, lower y value
        YIH-inside of search band, higher y value
        using the same procedure that was used for the 
        x coordinates.
    
    Calculate the Laplacian within the search band:
      For every Y from Y=2,N2-1 
        IF((XOH & XOL are defined).AND.(XIH & XIL are undefined))    
          Calculate the Laplacian between XOL & XOH.
        IF((XOH & XOL are defined).AND.(XIH & XIL are   defined))    
          Calculate the Laplacian between XOL & XIL
      For every X from X=2,N1-1 
        IF((YOH & YOL are defined).AND.(YIH & YIL are undefined))    
          Between YOL & YOH 
            if(the Laplacian is zero) calculate the Laplacian.
        IF((YOH & YOL are defined).AND.(YIH & YIL are   defined))    
          Between YOL & YIL and between YIH & YOH
            if(the Laplacian is zero) calculate the Laplacian.
    
    If the pixel aspect ratio is 1.0 (i.e., PIXLENX=PIXLENY),
      the Laplacian is calculated as
            P(X,Y)-(P(X-1,Y-1)+P(X  ,Y-1)+P(X+1,Y-1)
                   +P(X-1,Y  )           +P(X+1,Y  )
                   +P(X-1,Y+1)+P(X  ,Y+1)+P(X+1,Y+1))/8.0
    
    If the pixel aspect ratio is not 1.0 (i.e., PIXLENX!=PIXLENY),
      the Laplacian is calculated as
            P(X,Y)-(P(X-1,Y-1)+P(X+1,Y-1)+P(X-1,Y+1)+P(X+1,Y+1))/8.0
                  -(P(X,Y-1)+P(X,Y+1))*(2*DX**2-DY**2)/4/(DX**2+DY**2)
                  -(P(X-1,Y)+P(X+1,Y))*(2*DY**2-DX**2)/4/(DX**2+DY**2)
            where DX=PIXLENX and DY=PIXLENY
    
    Search the Laplacian for the X,Y positions of the zero-crossings.  
    To qualify as a zero-crossing, the absolute value of the 
    Laplacian adjacent to the zero-crossing must be greater than a 
    threshold that is set by the input parameter, THRESH.  
    
    If the input parameter, THRESH, is not provided, i.e. = 0, THRESH
    will be determined from the center line and center column of the
    Laplacian:
    THRESH=0.5*MAX(MAX(ABS(W(I,N2/2)),I=1,N1),
                   MAX(ABS(W(N1/2,I)),I=1,N2))
    where W is the array of Laplacian values.
    
    Using THRESH, the zero-crossings are determined:
            NZC - the no. of zero-crossings
            XZC,YZC - the X,Y locations of the zero-crossings
    
    The search proceeds from XOL to XIL then from XOH to XIH
    for each Y from 2 to N2-1.  Next, the y direction is
    searched from YOL to YIL then from YOH to YIH 
    for each X from 2 to N1-1.
    
    Let W(I) represent a segment of a row or column from the
    Laplacian array that is being searched for the zero-crossings,
    with (I) increasing toward the center of the solar disk.
    If a zero-crossing occurs between W(I) & W(I+1), the
    exact position is determined by linear interpolation.
    The zero-crossing is used if 
      |W(I-N)|>THRESH &
      |W(I-N)|>|W(I-N+1)|>...>|W(I)| &
        where N can be >= 0
      |W(I+1)|>0.75*THRESH &
      |W(I+2)|>0.75*THRESH
    
    There is an Upper and Lower Limit for the no. of zero-crossings. 
    
    Lower Limit Test:
            If the no. of zero-crossings is less than
                    0.75*(AE+MIN(XC,AE)+BE+MIN(YC,BE))
            either
                    THRESH is decreased to THRESH/10 
            or if THRESH has previously failed the upper limit test
                    THRESH=0.5*(THRESH+value of THRESH at 
                                  the time of the upper limit test)
            and the zero-crossings search is repeated.
    
    Upper Limit Test:
            If the no. of zero-crossings is greater than 4*(N1+N2),
            either
                    THRESH is increased to THRESH*10.0 
            or if THRESH has previously failed the lower limit test
                    THRESH=0.5*(THRESH+value of THRESH at 
                                  the time of the lower limit test)
            and the zero-crossings search is repeated.
    
    The adjustment of THRESH and the search for zero-crossings 
    will be repeated 10 times, after which the program will fail
    with an error message.
    
    The least squares fit determines the parameters 
            XC,YC,CXX,CXY,CYY
    for the ellipse
            CXX*(X-XC)**2-CXY*(X-XC)*(Y-YC)+CYY*(Y-YC)**2=1.0
    by minimizing
            E=SUM(
                 (CXX*(XZCA(I)-XC)**2-CXY*(XZCA(I)-XC)*(YZCA(I)-YC)
                 +CYY*(YZCA(I)-YC)**2-1.0)**2
                 for I=1,NZC)
    
    The equations to determine the parameters are obtained by setting
    the derivatives of E with respect to the parameters equal to zero.
      XC:  -CXX*DX+CXY*DY/2.0
      YC:  -CYY*DY+CXY*DX/2.0
      AE:  (-COSA**2*DX**2-2.0*SINA*COSA*DX*DY-SINA**2*DY**2)/AE**3
      BE:  (-SINA**2*DX**2+2.0*SINA*COSA*DX*DY-COSA**2*DY**2)/BE**3
      ANG: +CXY*DX**2/2.0-(CXX-CYY)*DX*DY-CXY*DY**2/2.0
    where DX=XZCA(I)-XC and DY=YZCA(I)-YC
    and ANG above is in radians.
    
    The matrix is preconditioned so that the equation solver will fit 5 
    parameters even if the limb is circular (i.e., ANG is undefined 
    and if not preconditioned the equation solver will terminate with a 
    divide by zero).  This preconditioning consists of adding 0.005 
    times the average of the diagonal matrix elements for XC, YC, AE, and 
    BE to the diagonal matrix element for ANG.
    
    Spurious points are then rejected.
    The location of the zero crossings are compared to the new ellipse.
    The points for which
            ABS(1.-R).le.PCFIT/100.0
    are retained, where
            R=CXX*(XZCA(I)-XC)**2-CXY*(XZCA(I)-XC)*(YZCA(I)-YC)
                 +CYY*(YZCA(I)-YC)**2
    
    The least squares fit is performed a second time with the shorter
    list of zero-crossings.
    
    The least squares fitting algorithm can fail for a variety of reasons.
    Often these failures are caused by a search zone that does not contain
    the limb of the SUN.  Two failure modes are trapped:  the maximum 
    number of interactions is 50, and the semi-major and semi-minor axes 
    determined from each interaction must be less than the maximum 
    dimension of the image, MAX(N1,N2).  If either of these two failure 
    modes are detected, FNDLMB will set the limb parameters to the values 
    of the input parameters and if UPDHDR=YES, will add the parameter
    and value, FILLED=NO, to the image header and will continue processing
    normally.
    
    
EXAMPLES
    
    
TIME REQUIREMENTS
    
    
BUGS
    
    
SEE ALSO
    centro
    lmbsta
    lmbcor

\end{verbatim}

\begin{list}{}{}
\item $\,$
\item $\,$
\item $\,$
\item $\,$
\end{list}
\subsection{VELFIT help page}
\label{App_VELFIT}

Below is a transcription of the text that appears at:  http://gong.nso.edu/data/iraf\_help/velfit.html as accessed on December 14th, 2015 \citep{citeVELFIT}.

\begin{verbatim}

VELFIT (Feb97)                         ephem                        VELFIT (Feb97)



NAME
    velfit -- Fits the average solar observer motion  to  the  ephemeris
    velocity.
    
    
USAGE
    velfit in_file aver_file
    
    
PARAMETERS
    
    in_file
        The filename of containing the list of input images.
    
    aver_file
        The  qacV  file  containing  the  average velocity table.  These
        files are produced by VMICAL and are available in the DSDS.
    
    sigma = 3
        The clip value for the sigma filter.
    
    passes = 2
        The number of passes through the sigma clip filter.
    
    device
        Graphics output device  for  the  qa  plot.   Setting  device  =
        "none" or "" suppresses output.
    
    verbose
        Verbose  output,  useful  for  tracking the results through each
        pass of the filter.
    
    
DESCRIPTION

    VELFIT is a procedure to correct the observed Earth -  Sun  relative
    velocity   using  a  linear  fit.   The  program  extracts  all  the
    information it needs to perform the fit from the image  headers  and
    the  qac file. Then it calculates a bias and scale factor to correct
    the images.  These numbers are applied  to  the  average  velocities
    according to the equation:
    
        Ve = s * Va - B
    
    Where  Ve  is  the ephemeris velocity, Va is the average velocity, s
    is the scale factor and B is the bias correction.
   
    All images whose hour angle  is  less  than  30  degrees  above  the
    horizon  are  ignored.   The  sigma  filter  is  used to isolate and
    remove anomalous images.   After  the  bias  and  scale  factor  are
    computed,  they  are  applied  to  the  average velocities and their
    results are compared to the  sigma  clip  value.   Values  that  are
    higher  than  the  clip  value  are  removed  before the next filter
    pass.  The filter will not alter the image headers, the only way  to
    know if an image was removed is using the verbose option.
   
    VELFIT  updates  the header keywords VEL_BIAS and VELSCALE.  It does
    not apply those numbers to the images.
   
    NOTE: The values for sigma  and  passes  should  be  left  at  their
    default.
   
   
   
EXAMPLES

    1. Normal operation where printing is suppressed.
   
         cl> files bbvzi960324* > inlist
   
         cl> velfit inlist bbqacV960324 device="" verbose-
   
    2. Direct the QA plot to a printer and give full verbose output.
   
         cl> velfit inlist bbqacV960324 device=stdplot verbose+
   
   
TIME REQUIREMENTS

    About 10 minutes on a Sparc 20/61.
   
   
BUGS
   
   
SEE ALSO
    vfit

\end{verbatim}

\end{document}